\documentclass[twocolumn,aps,notitlepage,showpacs,floatfix]{revtex4-1}
\usepackage{epstopdf}
\usepackage{graphicx}
\usepackage{dcolumn}
\usepackage{bm}
\usepackage{amsfonts}
\usepackage{amsmath}
\usepackage{mdframed}
\usepackage[percent]{overpic}
\usepackage[colorlinks=true,citecolor=blue]{hyperref}
\hypersetup{colorlinks=true,citecolor=blue,linkcolor=blue,urlcolor=blue}
\DeclareRobustCommand{\rchi}{{\mathpalette\irchi\relax}}
\newcommand{\irchi}[2]{\raisebox{\depth}{$#1\chi$}}

\begin{document}
\title{Probing quantum criticality using nonlinear Hall effect in a metallic Dirac system}
\date{\today}
\author{Habib Rostami}
\affiliation{Nordita, KTH Royal Institute of Technology and Stockholm University, Roslagstullsbacken 23, SE-106 91 Stockholm, Sweden}
\author{ Vladimir Juri\v ci\'c}
\affiliation{Nordita, KTH Royal Institute of Technology and Stockholm University, Roslagstullsbacken 23, SE-106 91 Stockholm, Sweden}
\begin{abstract}
Interaction driven symmetry breaking in a metallic (doped) Dirac system can manifest in the spontaneous gap generation at the nodal point buried below the Fermi level. Across this transition linear conductivity remains finite making its direct observation difficult in linear transport. We propose the nonlinear Hall effect as a direct probe of this transition when inversion symmetry is broken.
Specifically, for a two-dimensional Dirac material with a tilted low-energy dispersion, we first predict a transformation of the characteristic inter-band resonance peak into a non-Lorentzian form in the collisionless regime. Furthermore, we show that inversion-symmetry breaking quantum phase transition is controlled by an exotic tilt-dependent line of critical points. As this line is approached from the ordered side, the nonlinear Hall conductivity is suppressed owing to the scattering between the strongly coupled incoherent fermionic and bosonic excitations. Our results should motivate further studies of nonlinear responses in strongly interacting Dirac materials.
\end{abstract}
\date{\today}
\maketitle
\section{Introduction}  \label{Sec:Intro}

Nonlinear response functions are extremely sensitive to the structural symmetry of crystalline systems.
In particular, the second-order spectroscopy, such as second harmonic generation (SHG), probing the second order conductivity, is a powerful technique to characterize the crystalline orientation of a sample \cite{Shen_book}. The SHG is forbidden in the presence of spatial inversion symmetry, and can therefore play the role of an order parameter distinguishing the phases across the transition at which the spatial inversion symmetry is broken.
Furthermore, there have recently been growing interest in the nonlinear (second-order) Hall effect \cite{Deyo_arxiv_2009,Moore_prl_2010,Hosur_prb_2011,Sodemann_prl_2015} which, unlike the usual one, occurs in the presence of time-reversal symmetry in non-centrosymmetric (semi)metals featuring tilted Dirac fermions (tDFs) at low energies, such as single and few-layer WTe$_2$ \cite{Ma_nature_2018,Ma_np_2018,Kang_nm_2019}.
Nonlinear Hall effect amounts to the generation of a transverse current as a second-order response to a linearly polarized external electric field, and, as has been recently shown,
it is controlled by the Fermi surface average of Berry curvature derivative, the so called {\it Berry curvature dipole}~\cite{Sodemann_prl_2015,Zhang_prb_2018,You_prb_2018,Quereda_np_2018,Facio_prl_2018}.
Other phenomena, e.g. injection and anomalous photocurrent in Weyl semimetals are also interesting nonlinear
phenomena related to the Berry curvature dipole
\cite{Fernando_nc_2017,Ma_np_2017,Zhang_prb_2018,Rostami_prb_2018,Juan_arxiv_2019,Matsyshyn_arxiv_2019}.
\par
\begin{figure}[t]
\centering
\includegraphics[width=80mm]{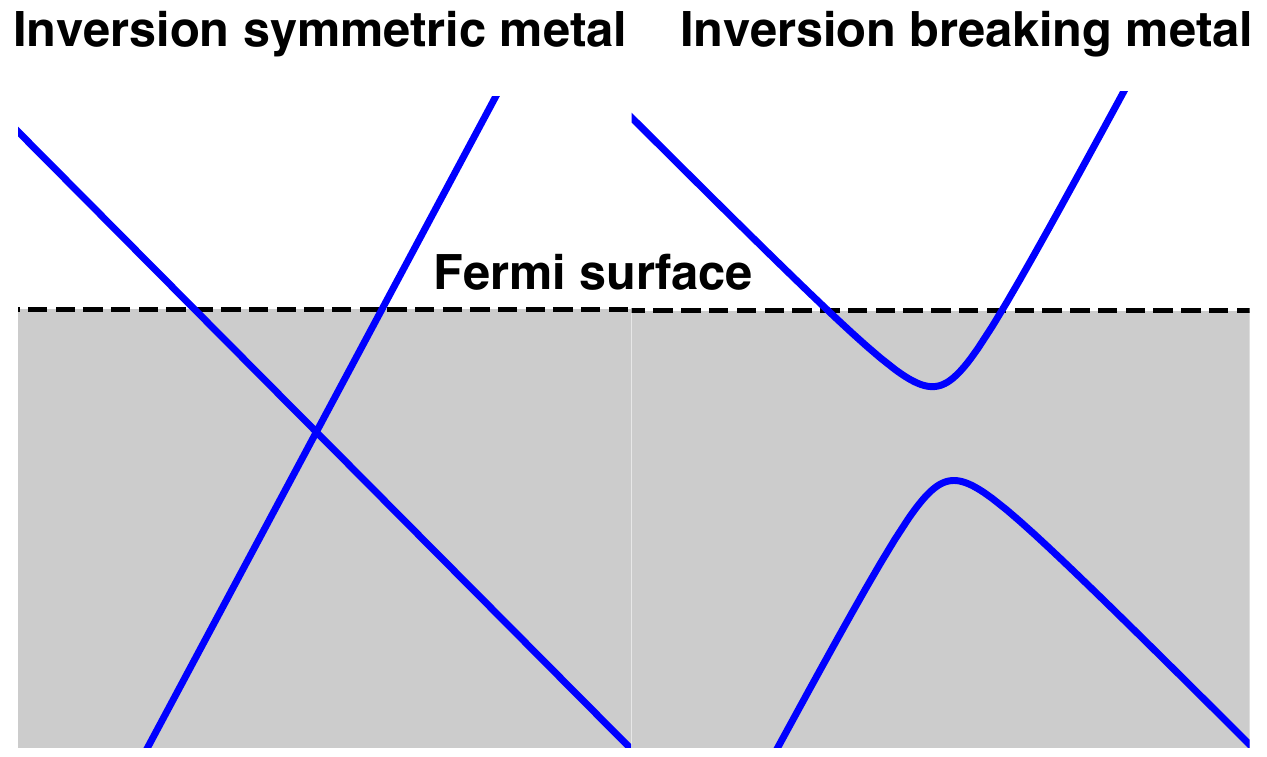}
\caption{Schematic plot for an inversion symmetric and inversion breaking semimetal described by tilted Dirac fermions (tDFs) at low energies. Although the gap opening occurs  away from the Fermi surface, it can have a strong impact on the nonlinear response of the system in the corresponding metallic phase ($\mu>0$).}
\label{fig:sketch}
\end{figure}

In this work, we show that the nonlinear Hall effect can be used as a powerful tool to probe the electron interaction driven inversion symmetry breaking in a metallic phase that emerges from a generic nodal band structure. In this case, the chemical potential is outside the gap region (see Fig.~\ref{fig:sketch}) and  the usual linear conductivity is finite in both symmetric and symmetry-broken phases. In contrast, the nonlinear conductivity is finite {\it only} when the gap is opened, i.e. the inversion symmetry is broken, and therefore may be used to directly detect the phase transition.

The inversion symmetry breaking at a finite chemical potential $\mu>0$  strongly relies on the presence of a Dirac point buried below the Fermi level. Namely, at a finite  chemical potential (but low enough so that the Dirac approximation is still valid), and at a strong short range (Hubbard-like) electron interaction, the band gap opening may occur at the Dirac point  because the system would optimally deplete the free energy as would so for $\mu=0$. The latter is expected since   quite generically there is no phase space for an interaction driven Mott insulating instability to take place at a finite but low enough $\mu$~\cite{Shankar_rmp_1994}. Irrespective of whether the system is electron-doped ($\mu>0$) or hole-doped ($\mu<0$), this scenario is expected to remain operative up to a critical chemical potential,  at which a superconducting instability takes over;   see also Ref.~\cite{Roy_Juricic_TBLG} for the discussion on the stability of a doped interacting Dirac liquid without the tilt.

We therefore consider strongly interacting two-dimensional  tDFs at the neutrality point ($\mu=0$) and at zero temperature ($T=0$) in the vicinity of a quantum phase transition (QPT) into an inversion-symmetry broken phase within the framework of the Gross-Neveu-Yukawa (GNY) quantum critical theory. By now it is well established that the GNY field theory captures the behavior of strongly interacting gapless Dirac fermions in the vicinity of a QPT to an ordered gapped phase ~\cite{Zinn_book,Rosenstein1993,HJV2009,Roy2011,RJH2013,Ponte_2014,RJ2014,Jian2015}. Furthermore, the field-theoretical predictions have been corroborated by the results from the numerical
(mostly quantum Monte Carlo) simulations ~\cite{
Sorella_srep_2012,Assaad_prx_2013,Parisen_prb_2015,Otsuka_prx_2016}. By employing a renormalization group (RG) analysis to the leading order in the $\epsilon~(=3-D)$ expansion close to the upper critical $D=3$ spatial dimensions,  we show that in the presence of the tilt
a QPT into an inversion-symmetry broken phase
 is controlled by an unusual tilt-dependent line of the quantum critical points (QCPs) which may be stable non-perturbatively, due to the pseudo-relativistic invariance of the boson-fermion Yukawa interaction and the specific form of the tilt term.

 We find that the dynamical nonlinear Hall conductivity (NLHC) is suppressed in the ordered (symmetry broken) phase close to this line of QCPs as compared to the noninteracting massive tDFs, and therefore can be used to probe such a symmetry breaking in a tDF metal. This effect can be traced back to the scattering of the strongly interacting soup of incoherent fermionic and bosonic excitations close to this line of QCPs. Furthermore, this suppression increases with the tilt parameter, consistent with the expectation based on the scaling of the density of states (DOS).

The paper is organized as follows: In Sec.~\ref{Sec:NLC-Scaling} we present a general scaling analysis of nonlinear conductivity in Dirac systems. In Sec.~\ref{Sec:NLC-NonInt} we analyze the main features of the second-order conductivity of two-dimensional tDFs in the collisionless regime  and zero temperature. In Sec.~\ref{Sec:GNY} we introduce the GNY quantum critical theory for tDFs, while in Sec.~\ref{Sec:LQCP} we perform the leading order renormalization group (RG) analysis of the theory and in Sec.~\ref{Sec:ICNLC} compute the interaction correction  to the nonlinear conductivity at the line of strongly coupled QCPs. Finally, materials aspects of our proposal were discussed in Sec.~\ref{Sec:Material}, and we conclude our work in Sec.~\ref{Sec:Discussion}.

\section{Nonlinear conductivity: Scaling analysis}\label{Sec:NLC-Scaling}

Dimensional analysis of the nonlinear conductivity in a $D$-dimensional Dirac system with linear dispersion (the dynamical exponent $z=1$) yields ${\rm dim}[\sigma^{(n)}] = D-2n$ in units of momentum (inverse length), as shown in Appendix \ref{app:dim}.
We only address the collisionless regime where frequency $\omega\gg T$,  which is governed by the particle-hole excitations created by the external electric field, since in this regime the conductivity displays universal features dictated exclusively by the dimensionality, the dispersion of the low energy quasiparticle excitations and the electron-electron interactions~\cite{sachdev_2011}. For the tDFs at finite temperature and frequency, the scaling form of the nonlinear optical conductivity reads (see Appendix \ref{app:dim})
\begin{align}\label{eq:NOC-large-Nf}
\sigma^{(n)}(n\omega) &\sim  \frac{1}{\omega^{2n-D}} f^{(n)}\left(\frac{\omega}{T},\frac{\mu}{T},\frac{m_f}{T},\alpha,\{g\}\right)
\end{align}
where $f^{(n)}(\{X\})$ is a universal scaling function of the dimensionless parameters $\{X\}$, $\mu$ and $m_f$ stand for the chemical potential and fermion mass, respectively, while $\alpha$ and $\{g\}$ represent the tilt parameter and dimensionless couplings. We here only focus on the high harmonic generation case for which all the frequencies are equal, and for the notational clarity we use  $\sigma^{(n)}(\omega_1=\omega,\omega_2=\omega,\dots,\omega_n=\omega)\equiv\sigma^{(n)}(n\omega)$.
\par
In the proximity of the line of strongly coupled QCPs, given by Eq.~(\ref{eq:GNY-QCP}), which, as we show, governs the behavior of the tDFs at the  QPT, the nonlinear conductivity picks up a correction given by
 \begin{align}\label{eq:sigma_ast}
 \sigma^{(n)}_\ast(n\omega) = Z^{n+1}_\Psi \sigma^{(n)}(n\omega).
 \end{align}
Here, $Z_\Psi$ is the renormalization factor for the fermionic field at this line of QCPs, which is directly related to the corresponding anomalous dimension, and $\sigma^{(n)}(n\omega)$ is the nonlinear conductivity for the noninteracting massive tDFs. Vertex corrections are absent due to the gauge invariance. The case $n=1$ and $\alpha=0$ corresponds to the linear conductivity of the untilted Dirac fermions~\cite{Roy_prl_2018}.  We show that for the $T=0$ SHG ($n=2$) the correction explicitly reads
\begin{align}
\sigma^{(2)}_\ast(2\omega) = \left\{1-\frac{3}{N_f(4-\alpha^2)} \right \} \sigma^{(2)}(2\omega)
\end{align}
to the leading order in the $\epsilon$ and $1/N_f$ expansions, with $\sigma^{(2)}(2\omega)$ as the $T=0$ second-order conductivity of the noninteracting system, given by Eq.~(\ref{eq:noninteracting-conductivity}). We notice that the conductivity is suppressed as compared to the noninteracting system due to the strong interactions of the fermionic and the order-parameter (bosonic) fluctuations close to the line of QCPs. The suppression is a monotonously increasing function of the tilt parameter which is consistent with the increase of the DOS at any finite energy and at a finite tilt with respect to the untilted case. Furthermore, we obtain universal inter-band features in the NLHC:
{\it non-Lorentzian} resonance peaks in the collisionless regime stemming from the anisotropic Fermi surface at the finite tilt, with the position and the linewidth strongly depending on the tilt parameter. We note that the strong tilt-dependence of the {\it linear} optical properties of tDFs were previously discussed \cite{Katayama_jpsj_2006,Nishine_jpsj_2010}.
\par
\section{Second-order conductivity of noninteracting tilted Dirac fermions}  \label{Sec:NLC-NonInt}
We consider an external homogeneous vector potential, ${\bf A}(t)$, with the corresponding electric field
 ${\bf E}(t)= -\partial_t  {\bf A}(t)$, as the driving field. The local second-order conductivity is obtained by utilizing the Kubo formula
\begin{align}
\sigma^{(2)}_{abc}(\omega_1,\omega_2 ) = -\frac{\rchi^{(2)}_{abc}(i\Omega_1,i\Omega_2)}{\omega_1\omega_2}
\Big|_{ i\Omega_m \to \omega_m+i\delta}.
\end{align}
Notice that $\delta \to 0^+$, $m=1,2$, and the second-order susceptibility given in terms of a three-point imaginary-time correlation function
$\chi^{(2)}_{abc}(\tau_1,\tau_2) =  \langle {\cal T} \hat j_c(-\tau_2) \hat j_b(-\tau_1) \hat j_a(0) \rangle$/2
 where ${\cal T}$ stands for the time-ordering operation, $\hat j_a = -e\sum_{\bf k}\hat \psi^\dagger_{\bf k} \partial_{k_a}\hat {\cal H}_{\bf k}\hat \psi_{\bf k}$. The paramagnetic current in the interaction picture $\hat j_a(\tau) = e^{-\tau \hat{\cal K}} \hat j_a e^{\tau \hat{\cal K}}$ with $\hat{\cal K}=\sum_{\bf k}\hat \psi^\dagger_{\bf k} (\hat {\cal H}_{\bf k}-\mu\hat I) \hat \psi_{\bf k}$, corresponds to the single-particle Hamiltonian ($\hbar=1$),
 \begin{align}\label{eq:free-fermion-hamiltonian}
 \hat{\cal H}({\bf k}) = \alpha \zeta k_x\hat I+v(\zeta k_x\hat \sigma_x+k_y\hat \sigma_y)+m_f\hat \sigma_z,
 \end{align}
where $\zeta=\pm$ stands for two fermion-flavors, $m_f$ represents the fermion mass due to the inversion symmetry breaking, ${\hat {\bm \sigma}}$ are the Pauli matrices, $\hat I$ is the $2\times2$ unity matrix, and the Fermi velocity $v=1$ hereafter.
\par
We first calculate NLHC, $\sigma^{(2)}_{yxx}$, in response to a linearly polarized electric field along the $x$-direction. In principle, there are $dc$ and $ac$ contributions which correspond to rectification and second harmonic effects, respectively.  We here focus on the latter, since we consider the collisionless regime.
The zero temperature NLHC of an electron-doped system is given in terms of the Berry curvature through its derivative (see Appendix \ref{app:nonlin_formal})
\begin{align}\label{eq:noninteracting-conductivity}
\sigma^{(2)}_{yxx}(2\omega) = \frac{ie^3 }{\omega}
\sum_{\bf k,\zeta}  n_{\rm F}(\varepsilon^c_{\bf k})
\frac{\partial  \Omega_{yx}(k) }{\partial k_x}
 ~C\left(\frac{\omega}{2\varepsilon_k}\right).
\end{align}
 Here, $n_{\rm F}(\varepsilon) =\Theta(\mu-\varepsilon)$ is the Fermi-Dirac distribution function at $T=0$, $\varepsilon^c_{\bf k}=\zeta \alpha k_x+ \varepsilon_k$ with $\varepsilon_k = \sqrt{k^2+m^2_f}$ is the conduction band dispersion, $\Omega_{yx}(k)= \zeta m_f/2\varepsilon^3_k$ stands for the Berry curvature \cite{Xiao_rmp_2009} and $C(x) = 1/[(1-4x^2)(1-x^2)]$.
We see that even though the system is metallic, there is a strong interband correction to the NLHC, characteristic for  the collisionless regime.
\begin{figure}[t!]
\centering
\includegraphics[width=90mm]{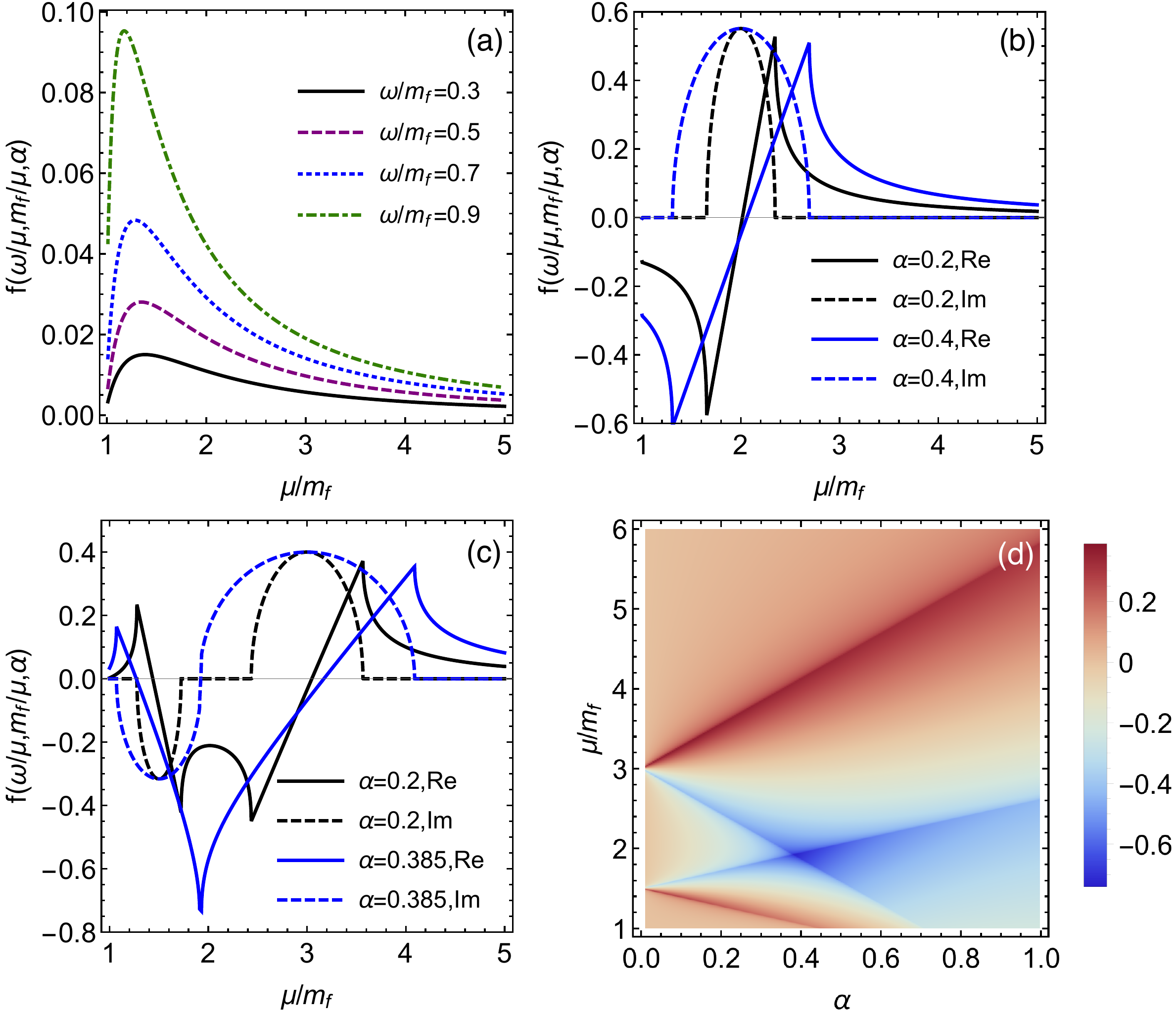}
\caption{Dependence of the universal function $f$, defined in Eq.~(\ref{eq:universal-f}),  with respect to $\mu/m_f$ and $\alpha$ for several values of $\omega/m_f$, as calculated in Appendix \ref{app:nonlin_f}. (a) The intraband regime for which $\omega<\mu$ and we set $\alpha=0.2$. The function $f$ is purely real in this regime. (b) An interband regime with $\omega=2m_f$. We can see that the position of peaks and dips strongly depends on $\alpha$ and an imaginary part in the resonance region appears.
(c) An interband regime with $\omega=3m_f$ for which there are two resonances. At a specific value of $\alpha \approx 0.385$ two dips merge. (d) Color plot for the real part of the function $f$ at $\omega=3m_f$ in which the shift in the position of peaks and dips can be traced. }
\label{fig:f_function}
\end{figure}
\par
The NLHC in the noninteracting system, after a momentum integration in Eq.~(\ref{eq:noninteracting-conductivity}),
explicitly reads
\begin{align}\label{eq:universal-f}
\sigma^{(2)}_{yxx}(2\omega)  =\frac{i e^3}{\omega^2} f\left( \frac{\omega}{\mu}, \frac{m_f}{\mu} ,\alpha\right)
\end{align}
where the universal function is to leading order in $\alpha$ given by (see Appendix \ref{app:nonlin_f})
 \begin{align}\label{eq:f-xyalpha}
f(x,y,\alpha) \approx \frac{12\alpha}{\pi^2}  \frac{xy (1-y^2)\Theta(1-y)}{(x^2-1)(x^2-4)}.
\end{align}
In the case of $\alpha\ll1$, the interband resonances occur at $x=\omega/\mu \approx 1 $ and $x=\omega/\mu \approx 2$. The exact dependence of $f(x,y,\alpha)$ on its arguments for $0\le\alpha\le1$ %arbitrary value of the tilt parameter ($\alpha\le1$)
is given by a quite lengthy analytical expression explicitly displayed in Appendix \ref{app:nonlin_f}, plotted  in Fig.~\ref{fig:f_function}.
\par
The corresponding result for the intraband regime, $\omega<\mu$, is depicted in Fig.~\ref{fig:f_function}(a) and the $f$-function is real-valued, similar as in Ref.~\cite{Sodemann_prl_2015}. Its form in the interband regime, $\omega>\mu$, is displayed in Fig.~\ref{fig:f_function}(b-d) where we can see that both the position and the shape of interband resonances strongly depend on the value of $\alpha$. This effect can be explained by considering the anisotropic Fermi surface which leads to a  momentum-dependent optical gap, $\Delta({\bf k})$, for the finite $\alpha$ case.
The single and multi-photon resonances occur when $n\omega$, with $n=1,2,\dots$, is equal to the optical gap at each momentum. Explicitly, in the presence of the tilt, the optical gap  $\Delta({\bf k}) = 2\sqrt{m^2_f+k^2}$ where ${\bf k}=(k_x,k_y)$ runs over the Fermi surface $(k_x+\zeta k_0)^2+k^2_y/(1-\alpha^2)=R^2$ in which $k_0=\alpha \mu /(1-\alpha^2)$ and $R=\sqrt{\mu^2-(1-\alpha^2)m^2_f}/\sqrt{1-\alpha^2}$. Such an optical gap at finite $\alpha$ leads to a splitting of the inter-band resonance peaks where the dip and the peak of the real part of the $f$-function shift in opposite directions. Simultaneously a broad {\it non-Lorentzian} resonance feature emerges in the imaginary part of the $f$-function (see Appendix \ref{app:nonlorenzian} for a detailed discussion). For $\alpha \ll1$, the Fermi surface is almost a circle with radius $k_{\rm F}$, which is the Fermi wave vector in the absence of the tilt. Therefore, the optical gap is nearly independent of the momentum $\Delta({\bf k})=2\sqrt{m^2_f+k^2_{\rm F}}=2\mu$. Accordingly, the interband resonances are quite sharp for the case of $\alpha\ll 1$.
Another nontrivial feature of the universal $f$-function is the cusp-like resonances (see Fig.~\ref{fig:f_function} (b) and (c)) in its real part steming from its logarithmic form, as explicitly given in Appendix \ref{app:nonlin_f}, see Eq.~(\ref{eq:f-func}). The corresponding one and two-photon resonances shift in the opposite directions in frequency by increasing the value of the tilt-parameter, $\alpha$. At a critical value of $\alpha$, the two resonances morph into a single one and then pass each other upon a further enhancement of $\alpha$, as can be seen in Fig.~\ref{fig:f_function} (d).

\section{Gross-Neveu-Yukawa quantum critical theory for tilted Dirac fermions}\label{Sec:GNY}
We now consider the effect of a  strong short range (Hubbardlike) electron interaction on the nonlinear optical conductivity within the framework of the GNY quantum critical theory for the tDFs.
The space-imaginary time action of the theory is  $S=S_f+S_Y+S_b$, and the non-interacting fermionic part reads
\begin{align}\label{eq:fermion-action}
S_f=\int d\tau\, d^D{\bf r}\,\,\psi^\dagger \left[\partial_\tau+\hat{\cal H}({{\bf k}\rightarrow-i\nabla_{\bf r}})\right]\psi
\end{align}
with the Dirac fermion field $\psi\equiv\psi(\tau,{\bf r})$, and  $\hat{\cal H}({\bf k})$ as the Hamiltonian for the noninteracting tDFs in Eq.~(\ref{eq:free-fermion-hamiltonian}) with $m_f=0$. The summation over the valley degree of freedom is assumed and we consider $2N_f$ copies of two-component Dirac spinors hereafter.
The short-range interaction is encoded through the Yukawa coupling between the Dirac fermion quasiparticles and the fluctuations of the underlying ordered state assumed to break Ising ($Z_2$) symmetry, as, for instance, a sublattice symmetry breaking charge density wave in graphene,
\begin{align}\label{eq:Yukawa}
S_Y= g \int d\tau\, d^D{\bf r}\,\, \phi \psi^\dagger{\hat\sigma}_z\psi.
\end{align}
Here, the bosonic field $\phi\equiv \phi(\tau,{\bf r})$ with the dynamics given by the action
\begin{align}\label{eq:boson}
S_b= \int d\tau\, d^D{\bf r} \left\{\phi (-\partial^2_\tau-\nabla^2+m_b^2) \phi+\frac{\lambda}{4!}\phi^4\right\},
\end{align}
where $m_b^2$ is the tuning parameter for the transition, and $m_b^2>0(<0)$ in the symmetric (symmetry broken) phase.

Engineering scaling dimensions of the Yukawa and $\phi^4$ couplings are ${\rm dim}[g^2]={\rm dim}[\lambda]=3-D$, while for the tilt parameter ${\rm dim}[\alpha]=0$, implying that $D=3$ is the upper critical dimension in the theory. We therefore use the deviation from the upper critical dimension as an expansion parameter $\epsilon=3-D$ to access the quantum-critical behavior in $D=2$. We set the bosonic and fermionic velocities to be equal to unity in the critical region~\cite{RJH-JHEP2016}.
\begin{figure}[t]
\centering
\includegraphics[width=80mm]{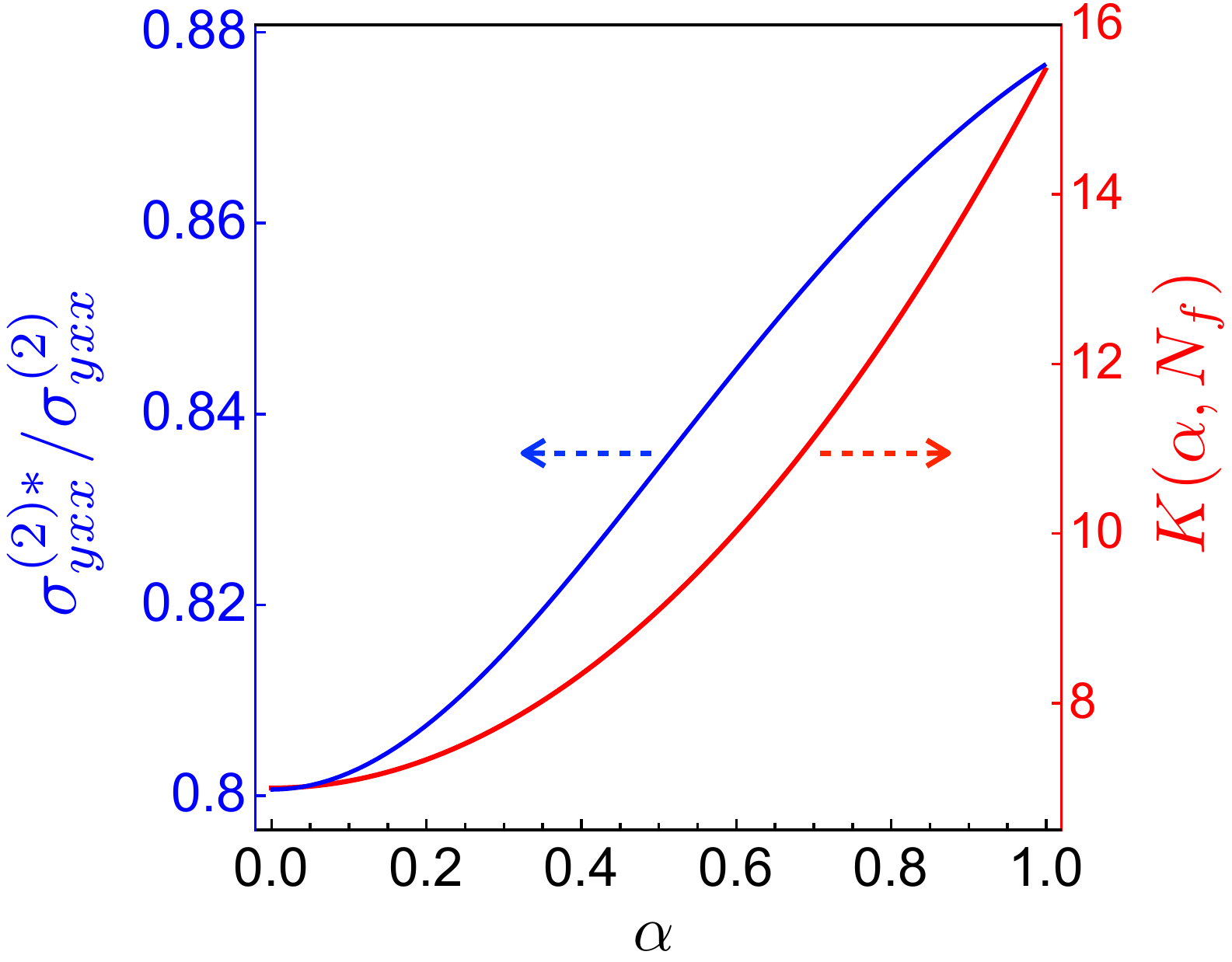}
\caption{The universal function $ K(\alpha,N_f)$ and the corrected NLHC relative to its bare value, $\sigma^{(2)\ast}_{yxx}/\sigma^{(2)}_{yxx}$, shown as a function of the tilt parameter, $\alpha$, in the spinless case, $N_f=2$.}
\label{fig:cond}
\end{figure}
\subsection{Renormalization group analysis: Line of quantum critical points}\label{Sec:LQCP}
To obtain the renormalization group (RG) flow of the couplings, we integrate out the modes with the Matsubara frequency $-\infty<\omega<\infty$, and then use the dimensional regularization in $D=3-\epsilon$ spatial dimensions within the minimal subtraction scheme. The RG $\beta$-functions to the leading order in the $\epsilon$-expansion, in the critical hyperplane ($m_b^2=0$, $m_f=0$), read (see Appendix \ref{app:GNY})
\begin{align}\label{eq:GNY-QCP}
&\beta_{g^2}=\epsilon g^2-g^4K(\alpha,N_f),\\
&\beta_{\lambda}=\epsilon \lambda-\frac{3}{2}\lambda^2-4N_f(1+2\alpha^2)\lambda g^2+24g^4 N_f,\\
&\beta_\alpha=0\label{eq:beta-alpha},
\end{align}
with $K(\alpha,N_f)$ calculated in Appendix \ref{app:GNY} and displayed in Fig.~\ref{fig:cond}, $\beta_Y\equiv -dY/d\ln(\kappa)$, $\kappa$ is the RG (momentum) scale, and the couplings are rescaled as $X/8\pi^2\to X$, with $X\in \{g^2,\lambda\}$. In the limit $\alpha\to0$ it can be readily seen that these $\beta-$functions reduce to the well known ones for the Ising GNY theory~\cite{HJV2009,RGJ2018}.
Crucially, for any $|\alpha|<1$, due to the marginality of the tilt parameter, the above flow equations yield a line of QCPs,  given by
\begin{align}
(g_\star^2,\lambda_\star)=\epsilon\left(\frac{1}{K(\alpha,N_f)},h(\alpha,N_f)\right),
\end{align}
with $h(\alpha,N_f)$ a complicated function of its arguments (see Appendix \ref{app:GNY}). Notice that  $K(\alpha,N_f)$ is a strictly positive and monotonously increasing function for $0\leq\alpha<1$, and therefore the value of the Yukawa coupling at the QCPs is smaller in comparison with untilted Dirac fermions;  the same holds for the $\phi^4$ coupling, $\lambda$.
This can be understood from the fact that the density of states (DOS) $\rho(E,\alpha)\sim |E|/\sqrt{1-\alpha^2}$, and therefore at any finite energy the DOS for tDFs increases as compared to untilted Dirac fermions until the system is over-tilted at $|\alpha|=1$. It is thus expected that the location of the critical point is pushed to a weaker coupling as the tilt increases and this feature is indeed captured in the RG analysis. However, the DOS is still vanishing at zero energy for a finite tilt implying that the critical points remain at a finite coupling, again consistent with the RG analysis. At this line of QCPs, both fermionic and bosonic excitations cease to exist as sharp quasiparticles because of their nontrivial anomalous dimension, respectively, given by $\eta_\Psi=g_\star^2 G(\alpha)/2$ and $\eta_\Phi=2N_f(1+2\alpha^2)g_\star^2$ with $G(\alpha)=4/(4-\alpha^2)$, obtained from $\eta_i=-(d Z_i/dg^2)\beta_{g^2}$ calculated at $X=X_\star$, with $Z_i(g^2)$, $i=\Psi,\Phi$, as the leading-order field renormalizations (see Appendix \ref{app:GNY}). Therefore, a family of non-Fermi liquids emerges from the QCP at a finite temperature for any $|\alpha|<1$.

We would like to emphasize that the marginality of the tilt parameter $\alpha$ as given by Eq.~\eqref{eq:beta-alpha} may be non-perturbative in nature, implying that the line of QCPs we found to the leading order in the $\epsilon-$expansion may be stable beyond this order. The reason for this lies in the pseudo-relativistic invariance of the Yukawa interaction [see Eq.~\eqref{eq:app-Yukawa} and the discussion therein], given in Eq.~\eqref{eq:Yukawa}, and the specific form of the tilt term. Namely, the tilt term manifestly breaks this symmetry and commutes with the rest of the free fermion action, as well as with the matrix entering the Yukawa term. Therefore, the Yukawa term is expected not to renormalize it, implying that the tilt parameter remains marginal.  On the other hand, it was shown that  manifestly Lorentz-symmetry breaking long-range Coulomb interaction renders the tilt parameter irrelevant~\cite{Lars2019}, consistent with the above argument.

\subsection{Interaction correction to the nonlinear conductivity at the line of quantum critical points}\label{Sec:ICNLC}
The leading order correction to the conductivity in the vicinity of the QCP is determined by the fermionic field renormalization $Z_\Psi$, as given by Eq.~(\ref{eq:sigma_ast}), which is ultimately related to the fermionic self-energy at vanishing external momentum, $\Sigma_f(i\Omega)$, explicitly evaluated in Appendix \ref{app:GNY}, yielding
\begin{align}
Z_\Psi = 1-  g_\star^2 \frac{1}{2\epsilon} G(\alpha).
\end{align}
Therefore, the nonlinear conductivity is modified according to Eq.~(\ref{eq:sigma_ast}) due to a strong  interaction between incoherent fermionic and bosonic excitations close to the QCPs. Explicitly,  the form of the correction to the NLHC reads
\begin{align}\label{eq:NLH-suppression}
\sigma^{(2)\ast}_{yxx}(2\omega) = \left(1-\frac{G(\alpha)}{2K(\alpha,N_f)}\right)^3 \sigma^{(2)}_{yxx}(2\omega),
\end{align}
which is displayed in Fig.~\ref{fig:cond}, and in the large $N_f$ limit yields the result shown in Eq.~(\ref{eq:NOC-large-Nf}).
%%%%%%%%%%%%%%%%%%%%%%%%%%%%%%%%%%%%%%%%%%%%%%%%%%%%%%%%%%%%%%%
 \section{Material realizations} \label{Sec:Material}
The case of WTe$_2$ is particularly interesting because of a very recent experimental observation of nonlinear Hall effect \cite{Ma_nature_2018,Ma_np_2018,Kang_nm_2019}. Actually, WTe$_2$ without the spin-orbit coupling  can be described in terms of  tilted massless Dirac fermions\cite{Ma_np_2018,Shi_prb_2019}, while the spin-orbit coupling opens up a direct gap located at $Q$ and $Q'$ valley points in the Brillouin zone \cite{Ma_np_2018,ZZDu_prl_2018} with the Berry curvature hot spots  localized around these points~\cite{Sodemann_prl_2015,Zhang_prb_2018,You_prb_2018,Quereda_np_2018,Facio_prl_2018}.
Considering the tilt parameter $\alpha   \approx 0.1$, the Fermi velocity $v \approx 0.5\times 10^6$m/s, the Dirac mass $m_f\approx 75$meV, the chemical potential $\mu \approx 1.3 m_f\sim 100$meV, and at $\hbar\omega\approx 120$meV \cite{Ma_nature_2018,Ma_np_2018,Kang_nm_2019}, we estimate the noninteracting NLHC to be $|\sigma^{(2)}(2\omega)| =  \sigma_0/E_0$ where $\sigma_0=e^2/\hbar$ and  $E_0 = (\hbar\omega)^2/(e \hbar v |f|) \approx 0.31$V/nm, with $|f|\approx 0.14$.
On the other hand, if this value of the mass gap is induced by the strong short range interaction close to the QCP, the NLHC should decrease as compared to this result for the noninteracting gapped (massive) tDFs, according to Eq.~(\ref{eq:NLH-suppression}).
\par
There are several other  candidates for the realization of massless tDFs in two dimensions such as 8Pmmn borophone \cite{Zhou_prl_2014,Mannix_Science_2015} with an electrically tunable tilt strength \cite{Farajollahpour_prb_2019},  topological crystalline insulators, such as SnTe \cite{Tanaka_np_2012} and organic compound such as $\alpha$-(BEDT-TTF)$_2$I$_3$ under pressure \cite{Kajita_jpsj_1992,Tajima_jpsj_2000,Kanoda-Science2017}.
Strong short range electron interactions, such as Hubbard onsite, may catalyze a mass gap therein, and the predicted behavior of the nonlinear Hall conductivity can be used to probe this phase transition. Furthermore, an analogue of twisted bilayer graphene featuring tilted and slow Dirac fermions at low energies may be an ideal candidate to realize the scenario we proposed in our work.   Finally, in three spatial dimensions, being the upper critical dimension for the GNY theory, only a correction to the conductivity stemming from the long-range Coulomb interaction, should remain~\cite{Roy_prl_2018,RJ2017}. This correction is expected to be $\alpha$-independent due to the irrelevance of the tilt parameter in this case \cite{Lars2019}.
\par
\section{Summary \& Outlook}\label{Sec:Discussion}
To summarize, we propose NLHC as an efficient tool to probe interaction tuned inversion symmetry breaking in the materials featuring Dirac fermions with the tilted dispersion. We show that the quantum critical behavior at a strong short-range interaction is governed by a line of QCPs from which a family of non-Fermi liquids emerges at a finite temperature. We find that NLHC decreases as the system approaches this line of QCPs from the ordered (symmetry broken) phase. Our results should motivate further studies of nonlinear response functions in strongly correlated Dirac materials, such as organic compound $\alpha$-(BEDT-TTF)$_2$I$_3$~\cite{Kanoda-Science2017}. Finally, the family of non-Fermi liquids we uncovered here should be further characterized in terms of optical and thermodynamic responses, which we plan to pursue in the future.
\par
\section*{Acknowledgements}
H.R. acknowledges the support from the Swedish Research Council (VR 2018-04252).
\bibliography{refs}
\pagebreak
\appendix

\begin{widetext}

\section{Dimensional analysis} \label{app:dim}
The light-matter interaction is defined by the following interaction term in the action:
\begin{align}
S_{int}  \sim  \int\,\,d^D x dt\,\, {\bf J}  \cdot {\bf A},
\end{align}
 where $D$ is the spatial dimension, ${\bf A}$ is the vector potential and the corresponding electric field reads ${\bf E} = -\partial_t {\bf A}$.
The current is given as the sum of linear and all nonlinear contributions
\begin{align}
{\bf J}\sim\sum_{n\ge1} {\bm \sigma}^{(n)} \underbrace{{\bf E}{\bf E}\dots{\bf E}}_{n\ \text{times}}.
\end{align}
Note that the dimension of the parameters in the units of momentum (inverse length) is
\begin{align}
{\rm dim}[S_{int}]=0~~,~~
{\rm dim}[x_\mu]=-1~~,~~
{\rm dim}[A_\mu]=1~~,~~
{\rm dim}[t] = -{\rm dim}[\omega] =-z~,
\end{align}
where $z$ stands for the dynamical exponent which implies an energy dispersion as $\varepsilon_k \sim k^z$. Therefore, we have
\begin{align}
{\rm dim}[E_\mu]= {\rm dim}[A_\mu]-{\rm dim}[t] = 1 +z
~~,~~
{\rm dim}[J_\mu] = D+z-1.
\end{align}
Eventually, we obtain
\begin{align}
{\rm dim}[\sigma^{(n)}] =  D-(n+1) -(n-1)z.
\end{align}
For the Dirac model we have $z=1$ which implies ${\rm dim}[\sigma^{(n)}] = D-2n$ and thus the scaling form of the $n^{\rm th}$ order conductivity reads
\begin{align}
\sigma^{(n)} \sim  \frac{1}{\omega^{2n-d}} f^{(n)} \left( \frac{\omega}{T},\frac{\mu}{T},\alpha,\{g\} \right).
\end{align}
Note that the linear conductivity of (undoped or intrisic) Dirac systems in $D=2$ is dimensionless while the second order one scales as $1/\omega^2$.
This implies that at low-frequency spectroscopy like GHz or even THz the nonlinear response can be considerably stronger than its linear counterpart and may be a very good instrument to probe those phenomena which are hidden within the linear response framework.
\section{Nonlinear Hall conductivity evaluated using the Kubo's formula}\label{app:nonlin_formal}
For the Dirac system with Hamiltonian linearly dependent on the momentum, such as the one we consider here for tDFs given in Eq.~\ref{eq:free-fermion-hamiltonian},
there is only one Feynman diagram for the second order susceptibility ($\rchi^{(2)}_{\alpha \beta\gamma}$), which is shown in Fig.~\ref{fig:chi2}, and is defined as
\begin{figure}[b]
 \begin{overpic}[width=40mm]{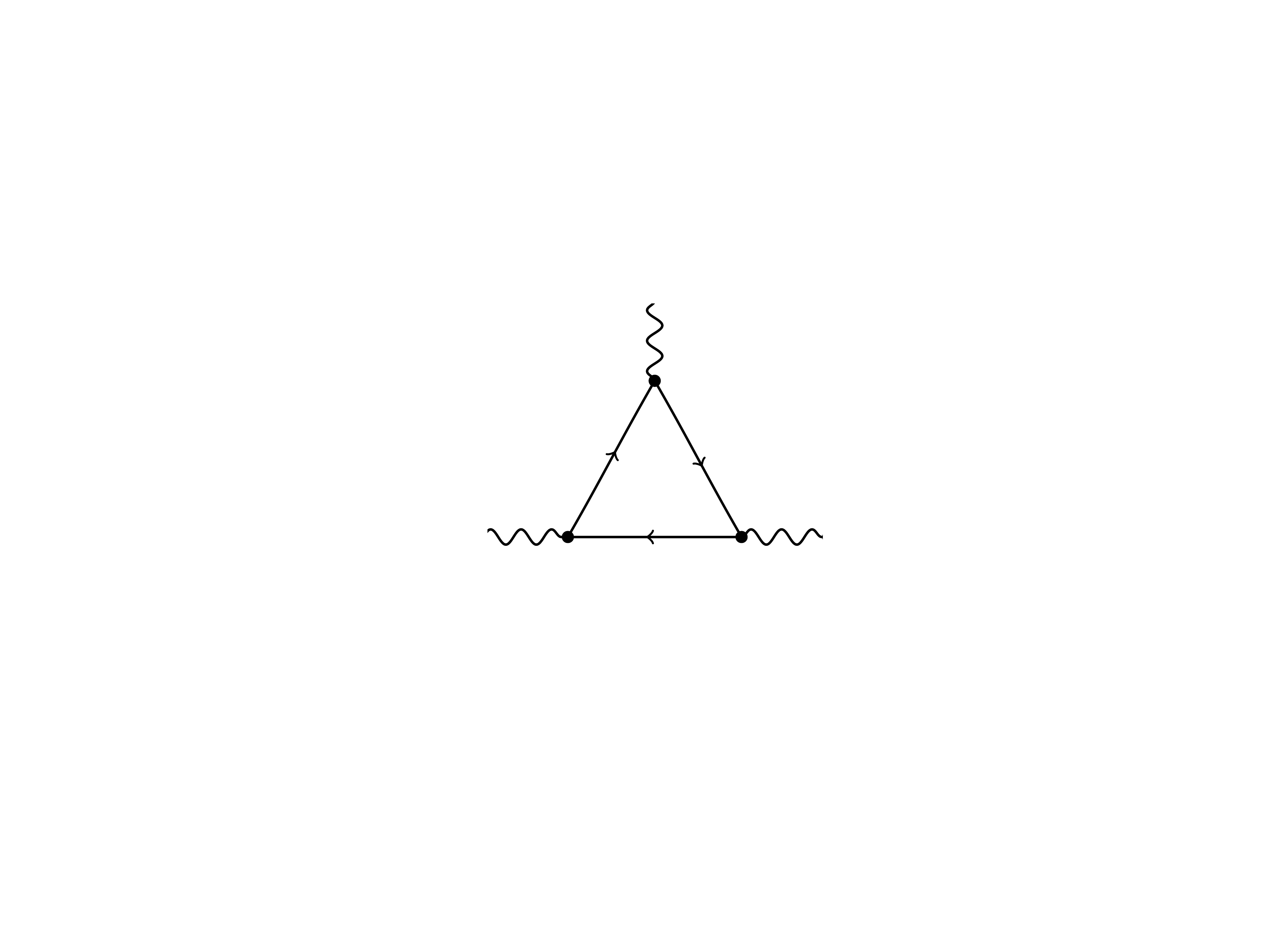}\end{overpic}
\caption{ Feynman diagram for the second order susceptibility which is a three-point correction function of paramagnetic current operator components. The solid line represents the fermionic propagator while the wavy line stands for the external photons.}
    \label{fig:chi2}%
\end{figure}
\begin{align}
J^{(2)}_\alpha = \sum_{\beta\gamma} \rchi^{(2)}_{\alpha \beta\gamma}  A_\beta A_\gamma.
\end{align}
The susceptibility corresponding to this bubble diagram reads \cite{Mahan}
\begin{align}
\rchi^{(2)}_{\alpha\beta\gamma}(m_1,m_2) =  \sum'_{\cal P} \sum_{\bf k}  \frac{1}{\beta}\sum_{n}  {\rm tr}
\Big[\hat \Gamma_\alpha  \hat G_f({\bf k},n) \hat \Gamma_\beta
\hat G_f({\bf k},n+m_1)\hat \Gamma_\gamma \hat G_f({\bf k},n+m_1+m_2) \Big],
\end{align}
where $\sum'_{\cal P}$ stands for the intrinsic permutation symmetry~\cite{Butcher_Cotter}, e.g. $(\beta,m_1)\leftrightarrow (\gamma,m_2)$,  and the fermionic Green's function is given by
\begin{align}
\hat G_f({\bf k},n) = \left(n \hat I-\hat{\cal H}({\bf k})\right)^{-1}.
\end{align}
Note that $\hat I$ is the identity matrix and $n$ ($m_i$) stands for the shorthand notation of the dummy (external) Matsubara frequency $i\omega_n= (2n+1)\pi/\beta$ ($i\omega_{m_i} = 2m_i \pi/\beta$) and $\beta=1/k_{\rm B}T$ where $k_{\rm B}$ the Boltzmann constant and $T$ is the temperature. The current vertex functions for the tDF Hamiltonian given in Eq.~\eqref{eq:free-fermion-hamiltonian} read
\begin{align}
\hat \Gamma_x = -e\partial_{k_x} \hat {\cal H}  = -e \zeta \hat\sigma_x -e\alpha \zeta\hat I
~~,~~
\hat \Gamma_y = -e\partial_{k_y} \hat {\cal H}  = -e \hat\sigma_y.
\end{align}
It is convenient to proceed with the calculation in the band basis which is denoted by $|{\bf k},\lambda\rangle$ with $\lambda\equiv \zeta,\ell$ where $\ell=c,v$ represent the conduction and valence band. We note that
\begin{align}
  \langle {\bf k},\lambda_1| \hat G({\bf k},n) |{\bf k},\lambda_2\rangle =
  \frac{\delta_{\lambda_1\lambda_2}}{n -\varepsilon^{\lambda_1}_{\bf k}}~~~,~~~
   \langle{\bf k}, \lambda_1|\hat \Gamma_\alpha |{\bf k},\lambda_2\rangle  = \Gamma^{\lambda_1\lambda_2}_\alpha({\bf k}),
\end{align}
where $\varepsilon^{\lambda=\pm}_{\bf k} = \zeta \alpha k_x +\lambda \varepsilon_k$  with $\varepsilon_k=\sqrt{k^2+m^2_f}$ as the energy eigenvalue for the conduction ($\lambda=+$) and valence ($\lambda=-$) band. We first perform the Matsubara summation over $n$ and subsequently an analytical continuation as $m_j \to \omega_j+i\delta$ with $\delta\to 0^+$. Eventually, we obtain
\begin{align}
\rchi^{(2)}_{\alpha\beta\gamma}(\omega_1,\omega_2) =  \sum'_{\cal P}    \sum_{\{\lambda_i\}}\sum_{\bf k}
  \frac{ \Gamma^{\lambda_1\lambda_2}_\alpha({\bf k})
 \Gamma^{\lambda_2\lambda_3}_\beta({\bf k})
 \Gamma^{\lambda_3\lambda_1}_\gamma({\bf k})
  }{\omega_1+\omega_2 +\varepsilon^{\lambda_2}_{\bf k}-\varepsilon^{\lambda_1}_{\bf k} +i\delta}
  \left\{ \frac{n_{\rm F}(\varepsilon^{\lambda_2}_{\bf k})-n_{\rm F}(\varepsilon^{\lambda_3}_{\bf k})}{\omega_1+\varepsilon^{\lambda_2}_{\bf k}-\varepsilon^{\lambda_3}_{\bf k} + i\delta}
  -
  \frac{n_{\rm F}(\varepsilon^{\lambda_3}_{\bf k})-n_{\rm F}(\varepsilon^{\lambda_1}_{\bf k})}{\omega_2+\varepsilon^{\lambda_3}_{\bf k}-\varepsilon^{\lambda_1}_{\bf k} + i\delta} \right\},
\end{align}
where $n_{\rm F}(x)=1/(e^{\beta(x-\mu)}+1)$ stands for the Fermi-Dirac distribution function.
From now on, we use a shorthand notation $\omega_j+i\delta \to \omega_j$.
We calculate $\rchi_{yxx}(\omega,\omega)$ as the {\it ac} nonlinear Hall response to an external driving field along $\hat {\bf x}$ direction. After summing over the band indexes we find
\begin{align}
\rchi^{(2)}_{yxx}(\omega,\omega) =
-\sum_{\bf k,\zeta}
\{n_{\rm F}(\varepsilon^c_{\bf k})-n_{\rm F}(\varepsilon^v_{\bf k})\}
 \left\{
  \frac{(\Gamma^{cc}_x-\Gamma^{vv}_x)\Gamma^{cv}_y\Gamma^{vc}_x}{(\varepsilon_{cv}-2\omega)(\varepsilon_{cv}-\omega)}
+
  \frac{(\Gamma^{cc}_x-\Gamma^{vv}_x)\Gamma^{cv}_x\Gamma^{vc}_y}{(\varepsilon_{cv}+2\omega)(\varepsilon_{cv}+\omega)}
+
  \frac{(\Gamma^{cc}_y-\Gamma^{vv}_y)|\Gamma^{cv}_x|^2 }{(\varepsilon_{cv}-\omega)(\varepsilon_{cv}+\omega)}
  \right\},
\end{align}
where $\varepsilon_{cv} =2\varepsilon_{k}= 2\sqrt{m^2+k^2}$.
We consider an electron doped case where the integral over the entire valence band is zero because the integrand is an odd function of $\bf k$. For an arbitrary function $f(x)$, we have $\int^{2\pi}_0 d\phi \sin\phi f(\cos\phi)
= 0$.  Using these two facts, we can make a further simplification:
\begin{align}
\rchi^{(2)}_{yxx}(\omega,\omega) &=- i\omega
 \sum_{\zeta,{\bf k}}
n_{\rm F}(\varepsilon^c_{\bf k})
 \frac{(\Gamma^{cc}_x-\Gamma^{vv}_x){\rm Im}[\Gamma^{cv}_y\Gamma^{vc}_x]}{\varepsilon^3_{cv}}
\frac{6 \varepsilon^4_{cv}}{\varepsilon_{cv}^4-5 \varepsilon_{cv}^2 \omega ^2+4 \omega ^4}.
\end{align}
We remind that the Berry curvature reads
\begin{align}
\Omega_{yx}({\bf k}) =-\frac{2}{e^2} \frac{{{\rm Im}[\Gamma^{cv}_y \Gamma^{vc}_x]}}{\varepsilon^2_{cv}} = \frac{\zeta m_f}{2(m^2_f+k^2)^{3/2}}.
\end{align}
Using the fact that $\Gamma^{cc}_x -\Gamma^{vv}_x = -e 2 \partial_{k_x} \varepsilon_{cv}$, we show that
\begin{align}
 \frac{ (\Gamma^{cc}_x -\Gamma^{vv}_x) {{\rm Im}[\Gamma^{cv}_y \Gamma^{vc}_x]}}{\varepsilon^3_{cv}}=
\frac{e^3}{6} \partial_{k_x} \Omega_{yx}(\bf k).
\end{align}
Eventually, the conductivity can be formally written as follows [see Eq.~(\ref{eq:noninteracting-conductivity})]
\begin{align}\label{eq:sigma2}
\sigma^{(2)}_{yxx}(2\omega) = -\frac{\rchi^{(2)}_{yxx}(\omega,\omega)}{\omega^2} = \frac{ie^3 }{\omega}
\sum_{\bf k,\zeta}
n_{\rm F}(\varepsilon^c_{\bf k})
\frac{\partial  \Omega_{yx}(k) }{\partial k_x}
 C\left(\frac{\omega}{2\varepsilon_{k}}\right),
\end{align}
in which
\begin{align}
C(x)
=\frac{1}{(1-4x^2)(1-x^2)}.
\end{align}
In the intraband regime $2\omega\ll m_f$ we have $x\to 0$ which leads to $C(x) \to 1$. Therefore, the inter-band contribution is related to the factor equal to $C(x)-1$.
\section{Analytical form of the nonlinear Hall conductivity}\label{app:nonlin_f}
In this section, we perform  the momentum integration in Eq.~(\ref{eq:sigma2}). The contribution from two valleys are equal which implies  $\sum_\zeta \to 2$ and we set $\zeta=1$. At zero temperature we have $n_{\rm F}(x)=\Theta(\mu-x)$ where $\Theta(x)$ is the Heaviside step function. Therefore, we have the following relation for the nonlinear Hall conductivity
\begin{align}
\sigma^{(2)}_{yxx}(2\omega) =2  \frac{ie^3}{\omega}
\sum_{\bf k}  \Theta(\mu-\alpha k\cos(\phi) - \varepsilon_k)
\frac{\partial  \Omega_{yx}(\bf k) }{\partial k_x}
 C\left(\frac{\omega}{2\varepsilon_k}\right).
\end{align}
Considering the derivative of the Berry curvature as $ {\partial  \Omega_{yx}(\bf k) }/{\partial k_x} =- 3m_f k\cos\phi /(2\varepsilon^5_k) $, we find
 \begin{align}
\sigma^{(2)}_{yxx}(2\omega) =-\frac{3}{(2\pi)^2}  \frac{ie^3}{\omega}
 \int^\infty_0 kdk \left[\int^{2\pi}_0 d\phi   \cos\phi \Theta(\mu-\alpha k\cos(\phi) - \varepsilon_k)\right]
\frac{m_f k}{\varepsilon^5_k} C\left(\frac{\omega}{2\varepsilon_k}\right).
\end{align}
 We then use the following identity that holds for any real $a,b$:
 \begin{align}
\int^{2\pi}_0 d\phi \cos(\phi) \Theta(a-b\cos(\phi)) = -2 {\rm sign}(b) \Theta\left (1+\frac{a}{b}\right) \Theta\left(1-\frac{a}{b}\right) \sqrt{1-\left(\frac{a}{b}\right)^2},
\end{align}
 to obtain
\begin{align}
\sigma^{(2)}_{yxx}(2\omega)
& = \frac{6}{(2\pi)^2} \frac{ie^3}{\omega}
 \int^\infty_0 kdk  \Theta(\alpha k-(\mu-\varepsilon_k) \Theta(\alpha k+\mu-\varepsilon_k)
 \sqrt{1-\left(\frac{\mu-\varepsilon_k}{\alpha k}\right)^2}
\frac{m_f k}{\varepsilon^5_k} C \left(\frac{\omega}{2\varepsilon_k}\right).
\end{align}
Now, we use another identity given below
\begin{align}
\int^{\infty}_0 dk~\Theta(\alpha k-(\mu-\varepsilon_k)) \Theta(\alpha k+\mu-\varepsilon_k) g(\alpha,k)
= \int^{k_1(\alpha) +k_0(\alpha) }_{k_1(\alpha) -k_0(\alpha) } dk~ g(\alpha,k ),
\end{align}
where
\begin{align}
k_0(\alpha) = \frac{\alpha\mu}{1-\alpha^2} ~~,~~k_1(\alpha) = \frac{\sqrt{\mu^2-m^2_f(1-\alpha^2)}}{1-\alpha^2}.
\end{align}
Note that for $\mu>m_f>0$ and $0<\alpha<1$ we always have $k_1(\alpha) >k_0(\alpha) >0$.
Therefore, we have
\begin{align}\label{eq:sigma2_2}
\sigma^{(2)}_{yxx}(2\omega) =   \frac{ie^3}{\omega^2}
f\left(\frac{\omega}{\mu},\frac{m_f}{\mu},\alpha \right),
\end{align}
where the universal function $f(\dots)$ is given by
\begin{align}
f\left (\frac{\omega}{\mu},\frac{m_f}{\mu},\alpha \right )= \frac{3 \omega m_f}{2\pi^2} \int^{k_1(\alpha)+k_0(\alpha) }_{k_1(\alpha)-k_0(\alpha) } dk~g(\alpha, k).
\end{align}
Here, we define
%
%\begin{align}
%S(\alpha,\mu,\omega,m_f) = \int^{k_1(\alpha)+k_0(\alpha) }_{k_1(\alpha)-k_0(\alpha) } dk~g(\alpha, k).
%\end{align}
%
 \begin{align}
 g(\alpha,k) = \frac{k/\alpha}{\varepsilon^5_k} \sqrt{(\alpha k)^2-(\mu-\varepsilon_k)^2 }
C\left(\frac{\omega}{2\varepsilon_k}\right).
 \end{align}
 \subsection{Solution of $f$-function for $\alpha \ll 1$}
 For a very small $1\gg \alpha>0$, we have
 %\begin{align}
%S=\int^{+k_0(\alpha)}_{-k_0(\alpha)} dk~g(\alpha,k_1(\alpha)+k) \approx 2 k_0(\alpha) g(\alpha,k_1(\alpha)).
% \end{align}
 %
 \begin{align}
f  = \frac{3 \omega m_f}{2\pi^2} \int^{k_1(\alpha)+k_0(\alpha) }_{k_1(\alpha)-k_0(\alpha) } dk~g(\alpha, k) \approx  \frac{3 \omega m_f}{2\pi^2}  \{2 k_0(\alpha) g(\alpha,k_1(\alpha))\}.
\end{align}
In this case,  we can also approximate
\begin{align}
k_0(\alpha) \approx \alpha \mu ~~,~~k_1(\alpha) \approx \sqrt{\mu^2-m^2_f}.
\end{align}
Accordingly, we arrive at the result given in Eq.~\eqref{eq:f-xyalpha} of the main text:
%
% \begin{align}
%S \approx2\alpha \frac{\mu^2-m^2_f}{\mu^4} C\left(\frac{\omega}{2\mu}\right)
%= 8 \alpha \frac{\mu^2-m^2_f}{(\mu^2-\omega^2)(4\mu^2-\omega^2)}.
% \end{align}
%
%
\begin{align}
 f  \approx \frac{12\alpha}{\pi^2}   \frac{\omega m_f(\mu^2-m^2_f)}{(\mu^2-\omega^2)(4\mu^2-\omega^2)}.
\end{align}
 \subsection{Solution of $f$-function for an arbitrary value of $\alpha\le 1$}
In this section we provide an exact solution for the $f$-function for arbitrary value of $\alpha\le 1$. We remind that
% \begin{align}
 % S=\int^{k_1(\alpha)+k_0(\alpha)}_{k_1(\alpha)-k_0(\alpha)} dk~g(\alpha,k).
 % \end{align}
%
 \begin{align}
f = \frac{3 \omega m_f}{2\pi^2} \int^{k_1(\alpha)+k_0(\alpha) }_{k_1(\alpha)-k_0(\alpha) } dk~g(\alpha, k)
\end{align}
 We introduce a new variable $y$:
 \begin{align}
 y= \sqrt{m^2_f+k^2} ~~,~~ ydy =kdk.
 \end{align}
Therefore, we have
%
%\begin{align}
%S= \frac{16}{\alpha} \int^{y_+(\alpha)}_{y_{-}(\alpha)}  dy
%\frac{ \sqrt{\alpha^2(y^2-m^2_f)-(\mu-y)^2 }}{(y^2-\omega^2)(4y^2-\omega^2)},
% \end{align}
 %
 \begin{align}
f= \frac{3 \omega m_f}{2\pi^2} \frac{16}{\alpha} \int^{y_+(\alpha)}_{y_{-}(\alpha)}  dy
\frac{ \sqrt{\alpha^2(y^2-m^2_f)-(\mu-y)^2 }}{(y^2-\omega^2)(4y^2-\omega^2)},
 \end{align}
 where $ y_{\pm}(\alpha) = \sqrt{m^2_f+(k_1(\alpha)\pm k_0(\alpha))^2}$.
It can be shown that $\alpha^2(y^2-m^2_f)-(\mu-y)^2 =(1-\alpha^2)\{a^2-(y-y_0)^2\}$
 in which we define
 \begin{align}
 a= \alpha k_1(\alpha)= \frac{\alpha\sqrt{\mu^2-(1-\alpha^2) m^2_f}}{(1-\alpha^2)}
 ~~,~~y_0= \frac{k_0(\alpha)}{\alpha}= \frac{ \mu}{1-\alpha^2}.
 \end{align}
After  a straightforward manipulation we find
%
% \begin{align}
%S=\frac{16\sqrt{1-\alpha^2}}{6\alpha \omega^3} \int^{y_{+}-y_0}_{y_{-}-y_0}  dy
%\sqrt{a^2-y^2 }
%\left\{\frac{1}{y+y_0-\omega}-\frac{1}{y+y_0+\omega}
%+\frac{2}{y+y_0+\omega/2}
%-
%\frac{2}{y+y_0-\omega/2}
%\right\}.
% \end{align}
 %
  \begin{align}
f=\frac{3 \omega m_f}{2\pi^2} \frac{16\sqrt{1-\alpha^2}}{6\alpha \omega^3} \int^{y_{+}-y_0}_{y_{-}-y_0}  dy
\sqrt{a^2-y^2 }
\left\{\frac{1}{y+y_0-\omega}-\frac{1}{y+y_0+\omega}
+\frac{2}{y+y_0+\omega/2}
-
\frac{2}{y+y_0-\omega/2}
\right\}.
 \end{align}
The explicit form of the above function is obtained after solving the master integral
 \begin{align}
 F(y,a,b) = \int dy \frac{\sqrt{a^2-y^2}}{y+b} =
 F_0(y,a,b) + F_1(y,a,b)
\end{align}
where
\begin{align}
 F_0(y,a,b) &= \sqrt{a^2-y^2}+b\arctan\left(\frac{y}{\sqrt{a^2-y^2}}\right)
 \\
 F_1(y,a,b) &=\sqrt{a^2-b^2}[\ln(b+y) -\ln\left(a^2+by+\sqrt{a^2-b^2}\sqrt{a^2-y^2}\right)].
 \end{align}
 Eventually, we obtain
 %
 % \begin{align}\label{eq:S-func}
%S = \frac{8\sqrt{1-\alpha^2}}{3\alpha \omega^3} \left\{Q(y_{+}-y_0) -Q(y_{-}-y_0) \right\}
% \end{align}
 %\begin{align}\label{eq:S-func}
%f = \frac{3 \omega m_f}{2\pi^2} \frac{8\sqrt{1-\alpha^2}}{3\alpha \omega^3} \left\{Q(y_{+}-y_0) -Q(y_{-}-y_0) \right\}
 %\end{align}
%
 \begin{align}\label{eq:f-func}
f =  \frac{4\sqrt{1-\alpha^2}}{ \pi^2 \alpha}    \frac{m_f[Q(y_{+}-y_0) -Q(y_{-}-y_0)]}{\omega^2}
 \end{align}
 where
 \begin{align}
Q(y)=   F_1(y,a,y_0-\omega)
 - F_1(y,a,y_0+\omega)
+2F_1(y,a,y_0+\omega/2)
-2F_1(y,a,y_0-\omega/2).
 \end{align}
It is worth to note that the contribution from $F_0(y,a,b)$ in the end cancels out in $Q$-function.
Eq.~(\ref{eq:f-func}) and Eq.~(\ref{eq:sigma2_2}) together yield the nonlinear conductivity in the bare bubble level. The numerical plots given in Fig.~\ref{fig:f_function}  are generated by using  Eq.~(\ref{eq:f-func}).
\section{Non-Lorentzian versus Lorentzian resonance}\label{app:nonlorenzian}
The nonlinear Hall conductivity is given in Eq.~(7) of the main text in which the universal function
$f(\omega/\mu,m_f/\mu,\alpha)$ shows interesting inter-band features. The inter-band resonance peak becomes broader when the value of the tilt parameter is increased while there is no scattering
mechanism in the model (we consider the collisionless regime).
Intriguingly, we can see the difference of this resonance with the usual Lorentzian one after
including a finite imaginary part for the frequency, e.g. $\omega\to \omega+i\Gamma$. The shape of the resonances for different values of $\alpha$ and $\Gamma$ are illustrated in Fig.~\ref{fig:f_image} and Fig.~\ref{fig:f_real}. For small $\alpha$ and finite $\Gamma$ the resonances feature a Lorentzian shape while by increasing the value of $\alpha $ they become broader without a considerable change in the peak value. The Lorentzian function changes its curvature sign from negative to positive when we move away form the resonance (see Fig.~\ref{fig:f_image}(c)) while in the non-Lorentzian case the curvature is always negative, see Fig.~\ref{fig:f_image}(a).
\begin{figure}[h]
 \centering
    \begin{overpic}[width=148mm]{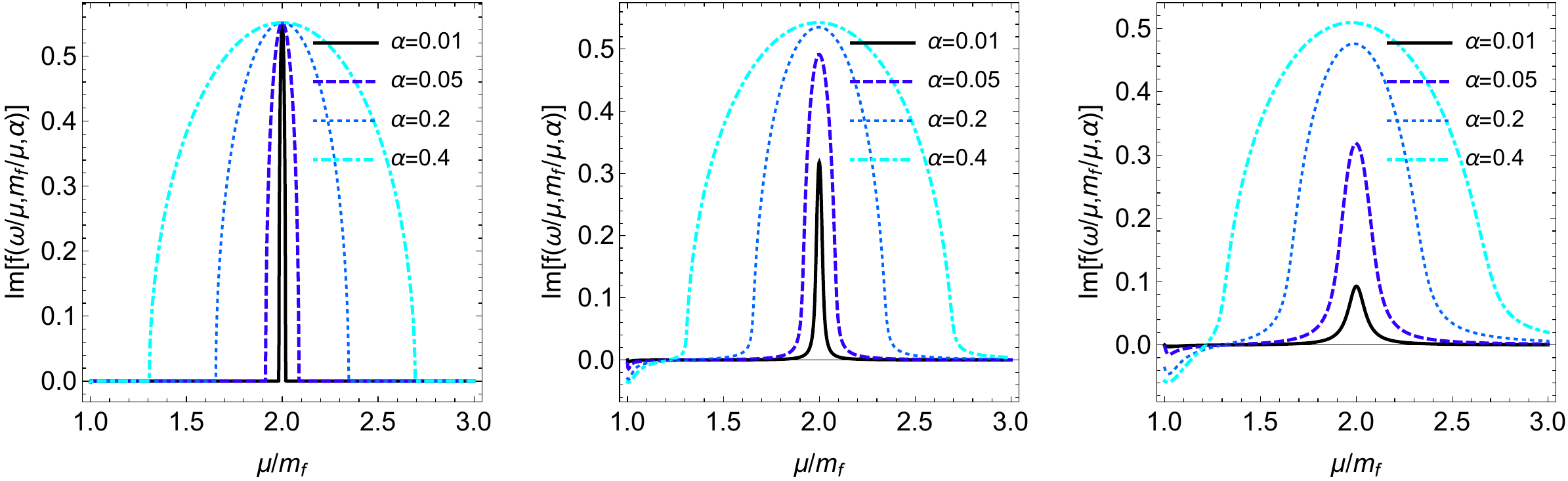}
    \put(7,28){(a)}
    \put(40.5,28){(b)}
    \put(74.5,28){(c)}
    \end{overpic}
\caption{Imaginary part of the universal $f$-function versus the chemical potential. We set $\omega=2m_f$ and clearly a resonance is seen at $\mu=\omega$. An imaginary part for the frequency is introduced as $\omega\to \omega+i\Gamma$.
We set $\Gamma\to 0^+$, $\Gamma=0.01m_f$ and $\Gamma=0.05m_f$ for (a),(b) and (c) panels, respectively.}
    \label{fig:f_image}%
\end{figure}
\begin{figure}[h]
 \centering
    \begin{overpic}[width=148mm]{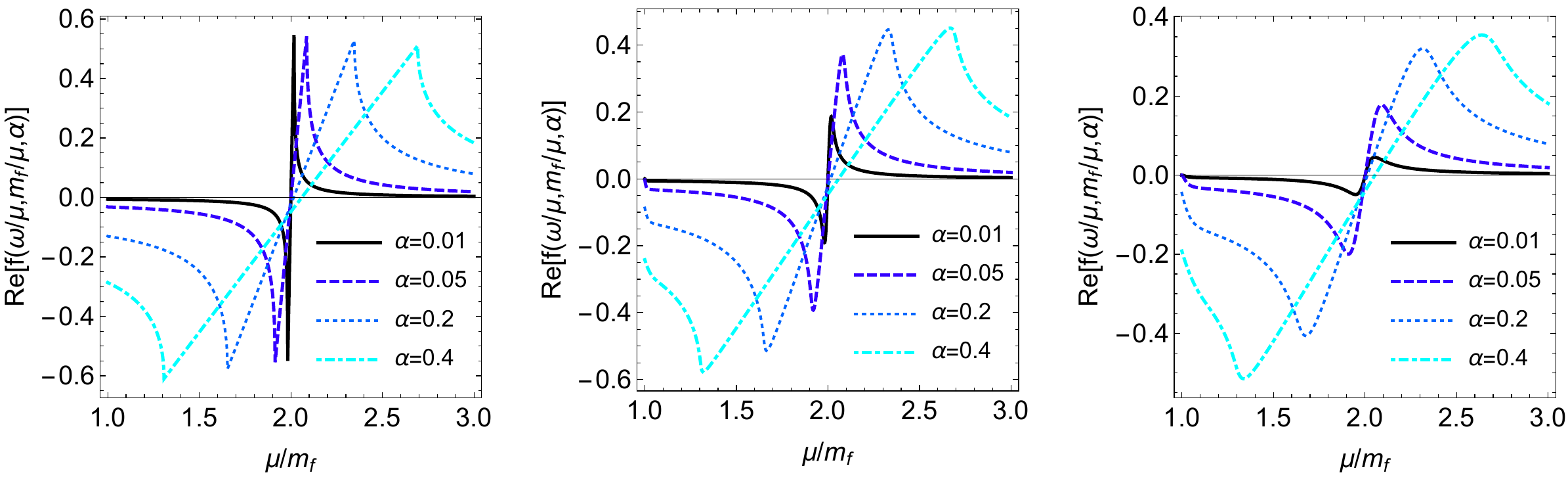}
    \put(8,27){(a)}
    \put(41.5,27){(b)}
    \put(75.5,27){(c)}
    \end{overpic}
\caption{Real part of universal $f$-function versus the chemical potential. We set $\omega=2m_f$ and clearly a resonance is seen at $\mu=\omega$. An imaginary part for the frequency is introduced as $\omega\to \omega+i\Gamma$.
We set $\Gamma\to 0^+$, $\Gamma=0.01m_f$ and $\Gamma=0.05m_f$ for (a),(b) and (c) panels, respectively. }
    \label{fig:f_real}%
\end{figure}
%
%\pagebreak
\section{Details of the renormalization group analysis of the Gross-Neveu-Yukawa theory for the tilted Dirac fermions}\label{app:GNY}
For the sake of completeness, we here repeat some of the details already presented at the beginning of Sec.~\ref{Sec:GNY}.
The momentum-imaginary time action for the noninteracting tilted Dirac fermions in $D$ spatial dimensions reads
\begin{align}
S_f=\int d\tau d^D{\bf r} \psi^\dagger(\tau,{\bf r})(\partial_\tau+ \hat{\cal H}({\bf k}\rightarrow -i\nabla_{\bf r}))\psi(\tau,{\bf r}).
\end{align}
Here, $\hat{\cal H}({\bf k})$ is the real space Hamiltonian for the free tilted massive Dirac fermions
obtained from its momentum space representation given by
\begin{align}
 \hat{\cal H}({\bf k})= \alpha \zeta k_x\hat I+v(\zeta k_x\hat \sigma_x+k_y\hat \sigma_y)+m_f\hat \sigma_z~.
 \end{align}
where $\zeta=\pm$ stands for two fermion flavors, $m_f$ represents the fermion mass due to the inversion symmetry breaking, ${\hat {\bm \sigma}}$ are the Pauli matrices, $\hat I$ is the $2\times2$ unity matrix, $|\alpha|<1$ is the tilt parameter, and the Fermi velocity $v=1$ hereafter.
The short-range interaction is encoded through the Yukawa coupling between the Dirac fermion quasiparticles and the bosonic degrees of freedom representing the fluctuations of the underlying ordered state, assumed to be a charge density wave for simplicity, and reads
\begin{align}\label{eq:app-Yukawa}
S_Y= g \int d\tau \int d^D{\bf r}\, \phi(\tau,{\bf r}) \psi^\dagger(\tau,{\bf r})\hat\sigma_z\psi(\tau,{\bf r}).
\end{align}
The dynamics of the bosonic fluctuations is described by the action
\begin{align}
S_b= \int d\tau \int d^D{\bf r} \{\phi(\tau,{\bf r}) (-\partial^2_\tau-\nabla^2+m_b^2) \phi(\tau,{\bf r})+\frac{\lambda}{4!}\phi^4\},
\end{align}
with $m_b^2$ the bosonic mass scaling as the distance from the  the QCP
$m_b^2$ is the tuning parameter for the QPT.
It can be readily shown that for each of the valleys $\zeta=\pm$, the above action can be casted in the manifestly Lorentz-invariant form, except for the tilt term, which explicitly breaks this symmetry. For instance, taking $\zeta=+$, to obtain such a form of the action, we can choose the $2\times2$ Dirac $\gamma$-matrices as $\gamma_0={\hat\sigma}_z, \gamma_1=-{\hat\sigma}_y, \gamma_2={\hat\sigma}_x$, and ${\bar\psi}=\psi^\dagger\gamma_0$. In particular, the Yukawa term in Eq.~\eqref{eq:app-Yukawa} can be rewritten in the relativistic form as $S_Y= g \int d\tau \int d^D{\bf r}\, \phi {\bar\psi}\psi$. Taking into account the other valley, the basis of the $4\times4$ $\gamma-$matrices can be chosen as $\Gamma_\mu=\{\Gamma_0, \Gamma_1,\Gamma_2\}=\{{\hat\sigma}_z\otimes{\hat\tau}_0, -{\hat\sigma}_y\otimes{\hat\tau}_z, {\hat\sigma}_x\otimes{\hat\tau}_0 \}$, with the ${\hat \tau}$ matrices acting in the valley space and ${\hat\tau}_0$ as the $2\times2$
unity matrix. In this basis the fermionic part of the action including now both (decoupled) valleys takes the Lorentz invariant form (except for the tilt term).

 Engineering scaling dimensions in the units of momentum (inverse length) of the Yukawa and $\phi^4$ couplings are ${\rm dim}[g^2]={\rm dim}[\lambda]=3-D$, while ${\rm dim}[\alpha]=0$, and therefore $D=3$ is the upper critical dimension in the theory. We will therefore use $\epsilon=3-D$, the deviation of from the upper critical dimension as an expansion parameter to access quantum-critical behavior in $D=2$.

\begin{figure}[h]%
\centering
\begin{overpic}[width=120mm]{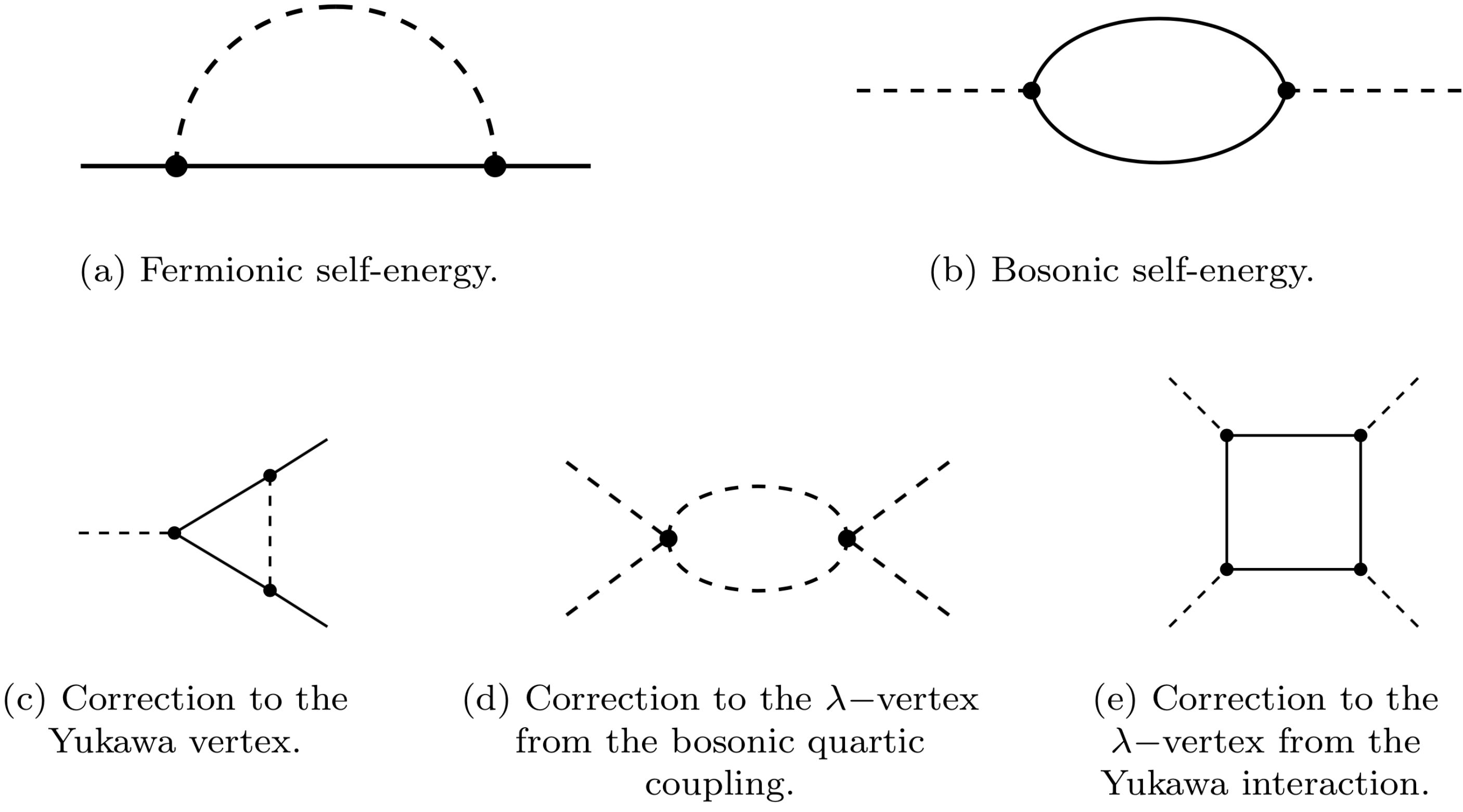}\end{overpic}
\caption{Feynman diagrams to the leading order in the $\epsilon$-expansion.
 The full line represents the fermionic propagator while the dashed line stands for the bosonic propagator.}
 \label{fig:FD}
\end{figure}

\subsection{Fermionic self-energy}

To find the correction to the nonlinear conductivity close to the QCP we compute the leading order zero-temperature ($T=0$) self-energy for the tilted Dirac fermions (see Fig.~\ref{fig:FD}(a))
\begin{align}
\Sigma_{f,\zeta}(i\Omega,{\bf k})=g^2\int [d\omega]\,\int [d^D {\bf q}]\, \hat\sigma_z\, G_{f,\zeta}(i\omega+i\Omega,{\bf k}+{\bf q})\hat\sigma_z
G_b(i\omega,{\bf q}),
\end{align}
where $[d\omega]\equiv d\omega/(2\pi)$, $[d^D{\bf k}]\equiv d^D{\bf k}/(2\pi)^D$, and the fermionic and bosonic propagators respectively read
\begin{align}
G_f(i\omega,{\bf k})&=\frac{i\omega+\alpha \zeta k_x+\tau \hat\sigma_x k_x+\hat\sigma_y k_y+m_f\hat\sigma_z}{(\omega+i \alpha\tau k_x)^2+k^2+m_f^2},\\
G_b(i\omega,{\bf k})&=\frac{1}{\omega^2+k^2+m_b^2},
\end{align}
and we set the bosonic and fermionic velocities to be equal to unity in the critical region~\cite{RJH-JHEP2016}.
After performing the matrix algebra in $D=2$, the above self energy in the critical hyperplane (all masses set to zero) explicitly reads
\begin{align}
\Sigma_{f,\zeta}(i\Omega,{\bf k})=g^2\int [d\omega] [d^D {\bf q}] \frac{i(\omega+\Omega)-\alpha \zeta (k_x+q_x)-\zeta \hat\sigma_ x(k_x+q_x)-\hat\sigma_y (k_y+q_y)}{\{[\omega+\Omega+i \alpha\zeta (k_x+q_x)]^2+({\bf k}+{\bf q})^2\}(\omega^2+q^2)}.
\end{align}

We now set ${\bf k}=0$ and $\zeta=1$ to obtain

\begin{align}\label{eq:FSE-1}
\Sigma_{f}(i\Omega,0)= - g^2\int \frac{d^D {\bf q}}{(2\pi)^D} \int^{\infty}_{-\infty} \frac{d\omega}{2\pi}
 \frac{i(\omega+\Omega)-{\bm \alpha}\cdot{\bf  q}-{\bm \sigma}\cdot{\bf q}}{\{[i(\omega+\Omega) - {\bm \alpha}\cdot {\bf q}]^2-q^2\}(\omega^2+q^2)},
\end{align}

with ${\bm \alpha}\equiv \alpha \hat{\bf x}$, and $\hat{\bf x}$ the unit vector in the $x$-direction. We perform the integral over $\omega$,

\begin{align}\label{eq:Sigma-f-inter}
\Sigma_{f}(i\Omega,0)=  - \frac{g^2}{2}\int \frac{d^D {\bf q}}{(2\pi)^D}
\frac{1}{q} \frac{i\Omega-{\bm \alpha} \cdot{\bf q}-{\bm \sigma}\cdot {\bf q}}{(i\Omega-{\bm \alpha}\cdot{\bf q})^2-4 q^2}.
\end{align}

Next, we drop the term proportional to $\bm \sigma\cdot\bf q$, since it can only generate a term of the form ${\bm\sigma}\cdot{\bm\alpha}$ which we eliminate by the corresponding counterterm, to find

\begin{align}
\Sigma_{f}(i\Omega,0)=  - \frac{g^2}{2}\int \frac{d^D {\bf q}}{(2\pi)^D}
\frac{1}{q} \frac{i\Omega-{\bm \alpha} \cdot{\bf q}}{(i\Omega-{\bm \alpha}\cdot{\bf q})^2-4 q^2},
\end{align}

which yields

\begin{align}
\lim_{i\Omega\to 0}\frac{\partial \Sigma_{f}(i\Omega,0)}{\partial (i\Omega)}=  \frac{g^2}{2}\int \frac{d^D {\bf q}}{(2\pi)^D}
\frac{1}{q} \frac{4q^2+({\bm \alpha} \cdot{\bf q})^2}{(4q^2-({\bm \alpha}\cdot{\bf q})^2)^2}.
\end{align}

We then integrate over the momentum using the hard cutoff $\Lambda$ in D=3, and use that $\bm \alpha = \alpha \hat {\bf x}$, to find
\begin{align}
\lim_{i\Omega\to 0}\frac{\partial \Sigma_{f}(i\Omega,0)}{\partial (i\Omega)}=  \frac{g^2}{(4\pi)^2} G(\alpha) \int^\Lambda_\lambda \frac{dq}{q}
\end{align}
with $\lambda=\Lambda/b$. Here, $b>1$ but $b-1\ll 0$ is the Wilsonian renormalization group (RG) parameter.
Finally, we use the correspondence between the hard cutoff and the dimensional regularizations
\begin{align}
\int^\Lambda_\lambda \frac{dq}{q}  \to \frac{1}{\epsilon},
\end{align}
which can be explicitly checked for instance by setting $\alpha=0$ in Eq.~(\ref{eq:FSE-1}), and keep only the divergent piece in the self energy. This procedure is systematically used in our analysis. Furthermore, we define
\begin{align}\label{eq:G-function}
G(\alpha) = 2\int^{2\pi}_0\frac{d\phi}{2\pi} \int^\pi_0 d\theta   \sin\theta
 \frac{4+\alpha^2(\sin\theta\cos\phi)^2}{(4-\alpha^2(\sin\theta\cos\phi)^2)^2}=4\int_0^\pi d\theta { \sin\theta}{(4 - \alpha^2 \sin^2\theta)^{-3/2}}=\frac{4}{4 - \alpha^2}=\frac{1}{1 - \alpha^2/4}.
\end{align}
Notice that this function is strictly positive for $|\alpha|<1$.

We then obtain the fermion field renormalization constant
\begin{align}\label{eq:Zpsi-final}
Z_\Psi = 1- \lim_{i\Omega\to 0} \frac{\partial \Sigma_{f}(i\Omega,0)}{\partial (i\Omega)}  =
1- \frac{g^2}{(4\pi)^2} \frac{1}{\epsilon} G(\alpha).
\end{align}

We now calculate $Z_\alpha$ using the condition
\begin{align}\label{eq:Zalpha-condition}
-Z_\Psi Z_\alpha \alpha k_x -k_x\frac{\partial\Sigma_f(i\Omega,{\bf k})}{\partial k_x}|_{i\Omega\to 0,{\bf k}\to 0}=-\alpha k_x.
\end{align}
We use Eq.~(\ref{eq:Sigma-f-inter}) for a finite external momentum ${\bf k}$, obtained by shifting ${\bf q}\to {\bf q}+{\bf k}$,
which then yields
\begin{align}
\frac{\partial\Sigma_f(i\Omega,{\bf k})}{\partial k_x}|_{i\Omega\to 0,{\bf k}\to 0}=\frac{g^2}{(4\pi)^2}\alpha G(\alpha)\frac{1}{\epsilon},
\end{align}
 with $G(\alpha)$ given by Eq.~(\ref{eq:G-function}).
Therefore, using this result, already obtained $Z_\Psi$ in Eq.~(\ref{eq:Zpsi-final}), and Eq.~(\ref{eq:Zalpha-condition}), we find that $Z_\alpha=1$, implying that the parameter $\alpha$ remains marginal to the leading order in the $\epsilon$ expansion.

\subsection{Self-energy for the bosonic field}

To find the RG flow equation for the Yukawa coupling we compute the self energy for the bosonic field and the correction to the Yukawa vertex.
We first compute the self energy for the bosons (Fig.~ \ref{fig:FD}(b)) which reads
\begin{align}
\Sigma_b(i\Omega,{\bf k})=-2g^2\int[d\omega]\int[d^D{\bf q}]\,{\rm tr}[\sigma_z\,G_f(i\omega,{\bf q})\sigma_z G_f(i\omega+i\Omega,{\bf q}+{\bf k})].
\end{align}
Since we are after the wave-function renormalization of the bosonic field, we set ${\bf k}=0$ in the above self energy.
After taking the trace, introducing the Feynman parameter, and shifting the integral over the frequency, we obtain
\begin{align}
\Sigma_b(i\Omega,{\bf k}=0)=4g^2 N_f\int_0^1\,dx\,\int[d\omega]\int[d^D{\bf q}]\,\frac{(\omega-x\Omega)(\omega+(1-x)\Omega)+q^2}{[\omega^2+(1-x)2i\alpha\Omega q_x+x(1-x)\Omega^2+q^2]^2},
\end{align}
with $2N_f$ as the number of two-component Dirac spinors.
The bosonic wave-function renormalization is determined by the following renormalization condition
\begin{align}
Z_\Phi-\frac{\partial\Sigma_b(i\Omega,0)}{\partial(\Omega^2)}|_{\Omega\to0}=1.
\end{align}
After taking the derivative and performing the remaining integrals, we obtain
\begin{align}
Z_\Phi=1-\frac{g^2}{8\pi^2}2N_f(1+2\alpha^2)\frac{1}{\epsilon}.
\end{align}

\subsection{$\beta$-function for the Yukawa coupling}

We now compute the correction to the Yukawa vertex (Fig.~\ref{fig:FD}(c); we take $\zeta=1$ for the valley index)
\begin{align}
\delta g=-g^3\int[d\omega]\int[d{\bf q}]\hat\sigma_z G_f(i\omega,{\bf q})\hat\sigma_z G_f(i\omega+i\Omega,{\bf q}+{\bf k})\hat\sigma_z G_b(i\omega,{\bf q}).
\end{align}
After setting external momentum and frequency to zero, since we are only after the divergent piece of the integral, we obtain
\begin{align}
\delta g=g^3\hat\sigma_z\int[d\omega]\int[d{\bf q}]\frac{1}{[(\omega+i\alpha q_x)^2+q^2](\omega^2+q^2)},
\end{align}
which, after carrying out the frequency and the momentum integrals, in turn yields
\begin{align}
\delta g=g^3\hat\sigma_z\frac{1}{8\pi^2}V(\alpha)\frac{1}{\epsilon},
\end{align}
where
\begin{align}
V(\alpha)=\frac{2}{|\alpha|}{\rm Arccosh}\left(\frac{2}{\sqrt{4-\alpha^2}}\right).
\end{align}
Renormalization condition for the Yukawa coupling reads
\begin{align}
Z_\Psi Z_\Phi^{1/2}g\hat\sigma_z -\delta g=g_0\kappa^{-\epsilon/2}\hat\sigma_z,
\end{align}
where $\kappa$ is the RG momentum scale \cite{Zinn-Justin}.
Using that the bare coupling $g_0$ is stationary under the RG transformation, we find
the infrared $\beta$ function for the Yukawa coupling $\beta_{g^2}\equiv-\frac{d g^2}{d\ln\kappa}$
\begin{align}\label{eq:beta-g}
\beta_{g^2}=\epsilon g^2-g^4[G(\alpha)+2V(\alpha)+2N_f(1+2\alpha^2)]\equiv \epsilon g^2-g^4 K(\alpha,N_f).
\end{align}
with
\begin{align}
K(\alpha,N_f)= G(\alpha)+2V(\alpha)+2N_f(1+2\alpha^2).
\end{align}
Here, we use that $\alpha$ is a marginal parameter to the leading order in the $\epsilon$-expansion and rescale $g^2/8\pi^2\to g^2$.
In the limit $\alpha\to0$ we obtain the known result for the $\beta$ function for the Yukawa coupling in the Ising GNY theory ~\cite{HJV2009,RGJ2018}
\begin{align}
\beta_{g^2}=\epsilon g^2-2g^4 \left( N_f+\frac{3}{2} \right).
\end{align}
For $|\alpha|<1$ different than zero, the location of the critical point
\begin{align}\label{eq:critical-g}
g_\star^2=\frac{\epsilon}{K(\alpha,N_f)}
\end{align}
is always pushed down in comparison to the case without the tilting $(\alpha=0)$, since the function $K(\alpha,N_f)$ defined in Eq.~(\ref{eq:beta-g}) is positive, even and monotonously increasing for $0<\alpha<1$.

\subsection{$\beta$-function for the $\lambda$ coupling}

Vertex correction of the $\lambda$ coupling for tilted Dirac fermions turns out to be equal to the case $\alpha=0$. Namely, the renormalization of the $\phi^4$ vertex does not depend on $\alpha$ as the bosonic propagator is $\alpha$-independent, see Fig.~\ref{fig:FD}(d), and reads
\begin{align}\label{eq:lambda-lambda}
\delta\lambda_\lambda=\frac{3}{2}\frac{\lambda^2}{8\pi^2}\frac{1}{\epsilon}.
\end{align}

Furthermore, the explicit calculation of the correction of the $\lambda$ coupling due to the Yukawa interaction (Fig.~\ref{fig:FD}(e)) gives the same result as without the tilt. This is a consequence of the fact that the tilt parameter can be eliminated in the corresponding integral by the shift of the frequency and the fact that the diagram is calculated in the limit of the vanishing external momentum and frequency. Explicitly, the divergent part of this contribution reads
\begin{align}\label{eq:lambda-Y}
\delta\lambda_Y=-\frac{24g^4}{4}\int[d\omega]\,\int[d^D{\bf q}]\, {\rm tr}[\hat\sigma_z G_f(i\omega,{\bf q})\hat\sigma_z G_f(i\omega,{\bf q})\hat\sigma_z G_f(i\omega,{\bf q})\hat\sigma_z G_f(i\omega,{\bf q})]=-24 N_f  \frac{g^4}{8\pi^2}\frac{1}{\epsilon}.
\end{align}

The renormalization condition of the $\lambda$ coupling reads
\begin{align}
Z_\Phi^{2}\lambda -(\delta \lambda_\lambda+\delta\lambda_Y)=\lambda_0\kappa^{-\epsilon},
\end{align}
with $\lambda_0$ as the bare coupling, and the term in the right hand side accounts for the scaling dimension of the $\lambda$ coupling ${\rm dim}[\lambda]=3-D=\epsilon$.
Using that the bare coupling is stationary under the RG transformation, and the results in Eqs.~(\ref{eq:lambda-lambda}) and (\ref{eq:lambda-Y}), we obtain the infrared $\beta$ function for the $\phi^4$ coupling
\begin{align}\label{eq:beta-lambda}
\beta_{\lambda}=\epsilon \lambda-\frac{3}{2}\lambda^2-4N_f(1+2\alpha^2)\lambda g^2+24g^4 N_f,
\end{align}
which in the limit $\alpha=0$ agrees with the known result for the $Z_2$ GNY theory for the untilted Dirac fermions~\cite{HJV2009,RGJ2018}.
Substituting the value of the critical Yukawa coupling given by Eq.~(\ref{eq:critical-g}) into the above $\beta$ function for the $\lambda$ coupling, we find that the fixed point is located at $\lambda_\star=\epsilon h(\alpha,N_f)$, with $h(\alpha,N_f)$ a complicated function of its arguments, plotted in Fig.~\ref{fig:beta-lambda}(a) as a function of the tilt  parameter for different values of $N_f$. The value of $\lambda_\star$ for $\alpha\neq0$ is smaller than that for $\alpha=0$, as expected from the scaling of the density of states for the tDFs.
The fixed point $X_\star\equiv(g_\star^2,\lambda_\star)$ is in fact a critical point since the eigenvalues of the stability matrix $M_{ij}(X)=\partial\beta_{X_i}/\partial X_j$, where  $X\equiv\{g^2,\lambda\}$, equal to its diagonal elements, are both negative at $X=X_\star$.  The eigenvaue $M_{11}(X_\star)=-\epsilon K(\alpha,N_f)<0$, since the function $K(\alpha,N_f)$ is positive for $|\alpha|<1$. The other eigenvalue  $M_{22}(X_\star)$  is also negative, as can be seen in Fig.~\ref{fig:beta-lambda}(b).

\begin{figure}
 \centering
    \begin{overpic}[width=50mm]{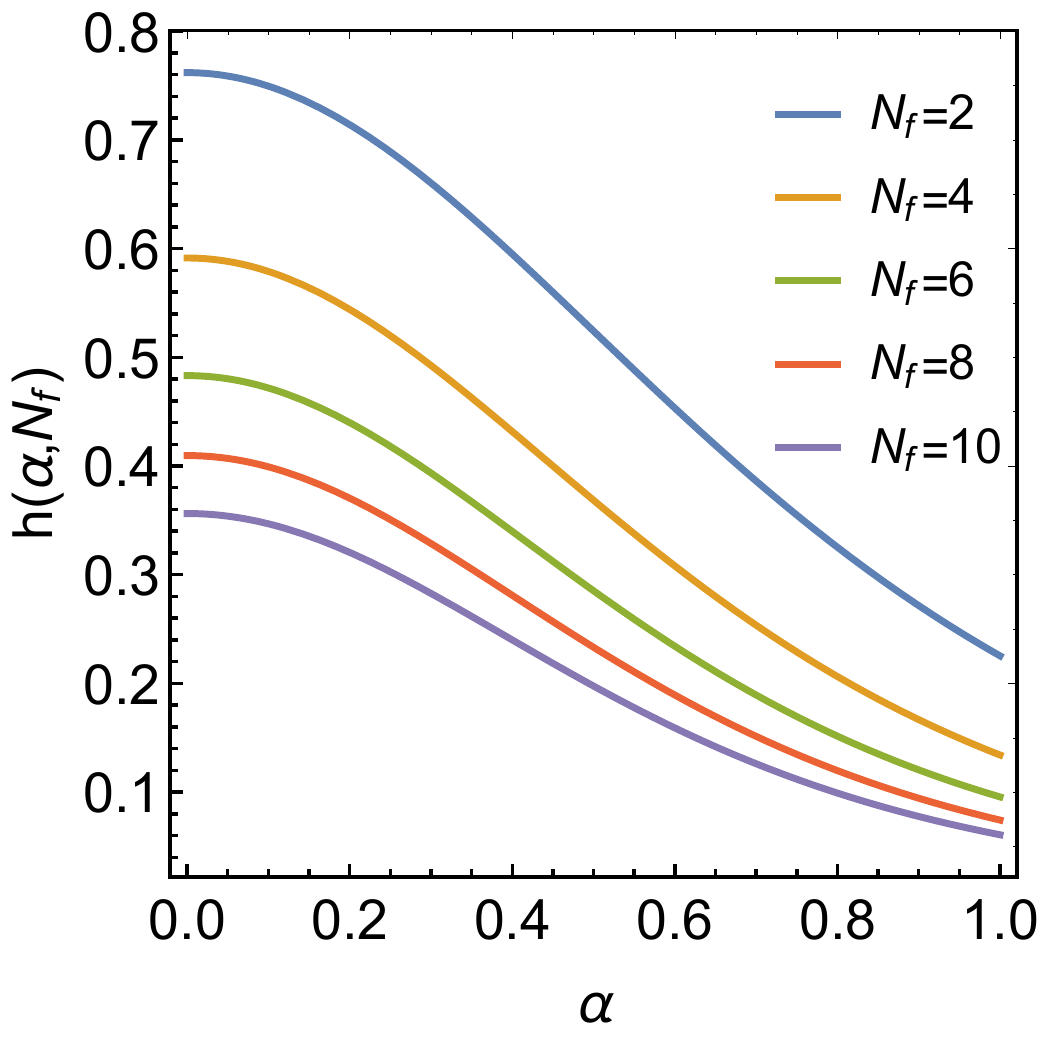}\put(25,80){(a)}\end{overpic}
    \begin{overpic}[width=52mm]{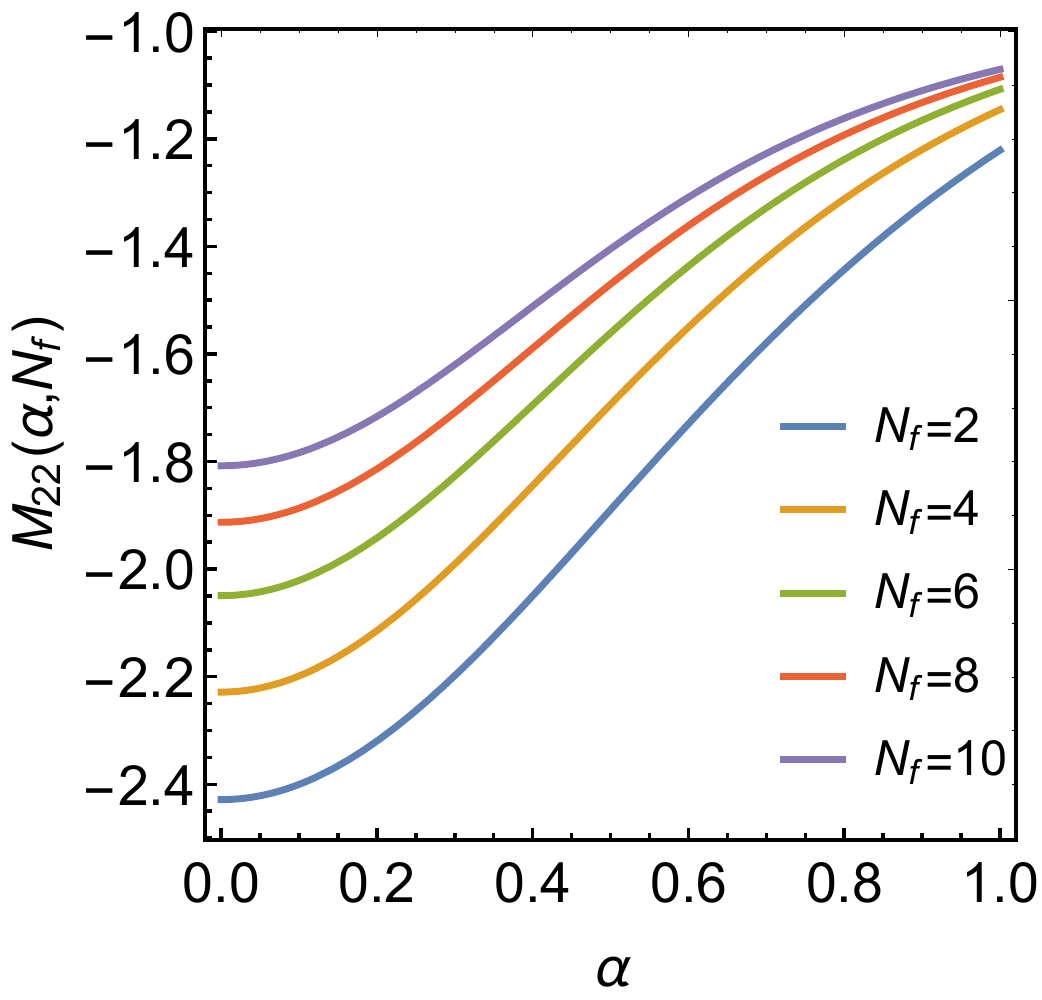}\put(25,80){(b)}\end{overpic}
\caption{(a) The function $h(\alpha,N_f)$ determining the value of the coupling $\lambda$ at the fixed point $X_\star=\{g^2_\star,\lambda_\star\}$, $\lambda_\star=\epsilon\, h(\alpha,N_f)$, of the RG flow given by Eqs.~(\ref{eq:beta-g}) and (\ref{eq:beta-lambda}), with $g^2_\star$ given in the text. (b) The eigenvalue $M_{22}$ of the stability matrix $M_{ij}(X)$ at the fixed point $X=X_\star$ showing that indeed the $\lambda$-direction is stable and therefore $X_\star$ is a QCP at least for the values of $N_f\leq10$.  }%
    \label{fig:beta-lambda}%
\end{figure}
\pagebreak
\end{widetext}

\end{document}